\documentclass[%
 superscriptaddress,
 amsmath,amssymb,
 prx, nofootinbib,
 twocolumn
]{revtex4-2}
\usepackage{graphicx} 
\usepackage{comment}
\usepackage{dsfont}
\usepackage{xcolor}
\usepackage{bm} 
\usepackage{hyperref}
\usepackage{color,soul}

\usepackage[utf8]{inputenc}
\usepackage{amsmath}
\usepackage{amssymb}
\usepackage[a4paper, total={6.6in, 8.8in}]{geometry}
\usepackage{graphicx}
\usepackage{hyperref}
\usepackage{float}
\usepackage{braket}
\usepackage{adjustbox}
\usepackage{color}
\usepackage{framed} 
\usepackage{url}
\usepackage[normalem]{ulem}

\usepackage{color}
\definecolor{eggplant}{RGB}{180,33,147}

\begin{abstract}
We report an accurate and efficient classical simulation of a kicked Ising quantum system on the heavy-hexagon lattice. A simulation of this system was recently performed on a 127 qubit quantum processor using noise mitigation techniques to enhance accuracy (Nature volume 618, p.~500–505 (2023)). Here we show that, by adopting a tensor network approach that reflects the geometry of the lattice and is approximately contracted using belief propagation, we can perform a classical simulation that is significantly more accurate and precise than the results obtained from the quantum processor and many other classical methods. We quantify the tree-like correlations of the wavefunction in order to explain the accuracy of our belief propagation-based approach. We also show how our method allows us to perform simulations of the system to long times in the thermodynamic limit, corresponding to a quantum computer with an infinite number of qubits. Our tensor network approach has broader applications for simulating the dynamics of quantum systems with tree-like correlations.
\end{abstract}

\begin{document}

\title{Efficient tensor network simulation of IBM's Eagle kicked Ising experiment}

\author{Joseph Tindall}
\affiliation{Center for Computational Quantum Physics, Flatiron Institute, New York, New York 10010, USA}

\author{Matthew Fishman}
\affiliation{Center for Computational Quantum Physics, Flatiron Institute, New York, New York 10010, USA}

\author{E. Miles Stoudenmire}
\affiliation{Center for Computational Quantum Physics, Flatiron Institute, New York, New York 10010, USA}

\author{Dries Sels}
\affiliation{Center for Computational Quantum Physics, Flatiron Institute, New York, New York 10010, USA}
\affiliation{Center for Quantum Phenomena, Department of Physics, New York University, 726 Broadway, New York, NY, 10003, USA}

\maketitle

\section*{Main}

\emph{Introduction.} Quantum computers are fundamentally noisy in nature and incur errors with each operation. To remedy noise and make the computer run as desired, one option could be quantum error correcting codes. In spite of significant advances in quantum technologies over the last decade, this type of error correction is currently out of reach. As such, huge efforts have been devoted to finding out whether current noisy quantum technologies could provide a practical advantage over classical computers without error correction. The emphasis here is on practicality, since expectation values of local observables in large classes of shallow quantum circuits can formally be computed in polynomial time on a classical computer~\cite{bravyi21, wild23} and so can the output distribution of a noisy random quantum circuit~\cite{aharonov23}. The latter refutes the claim of violation of the extended Church-Turing thesis in Ref.~\cite{arute19}, and makes it exceedingly unlikely any such violation can be achieved through noisy quantum computation without error correction. 
Noise can also have a similar effect as limiting entanglement, allowing noisy quantum computers to be simulated by
 tensor network methods \cite{Zhou2020, Pan2022, Ayral2023}.
While such works put constraints on the power of noisy quantum devices, they also show that it may be difficult to classically simulate them in practice when the noise is sufficiently small. At the same time, error mitigation techniques have been proposed to further improve the accuracy of noisy quantum devices by classical post processing~\cite{temme17,vandenberg23}. With these techniques one can construct a better classical (biased) estimator for the desired noiseless result by paying the overhead of having to run several noisy quantum circuits.

\begin{figure*}[t!]
    \centering
    \includegraphics[width = \textwidth]{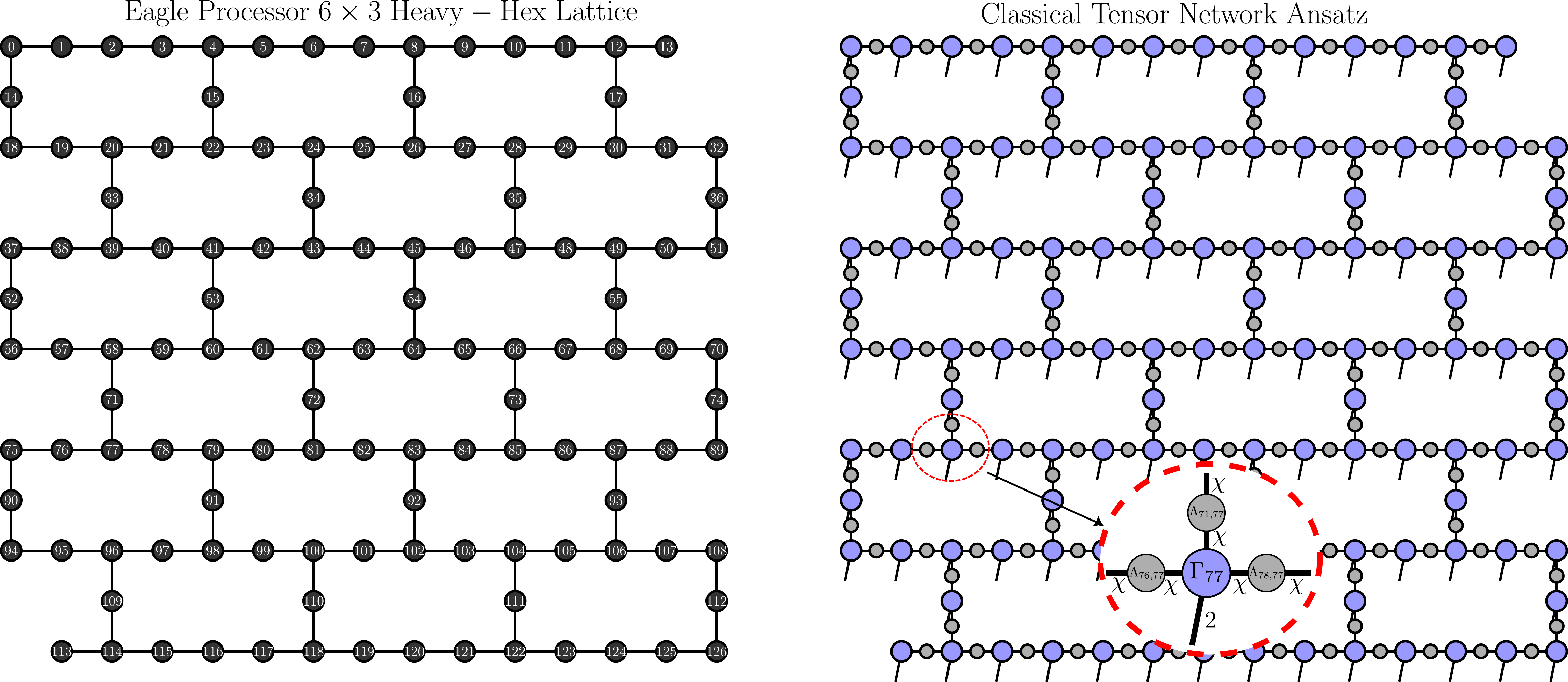}
    \caption{Left: Structure of the Eagle quantum processor which consists of a $6 \times 3$ heavy-hexagon lattice with two additional qubits ($113$ and $13$) added to the bottom left and right corners of the lattice. Right: Tensor network structure used for our simulations of heavy-hex lattices, with the network structure directly reflecting the lattice. On-site tensors $\Gamma_{v}$ are coloured in blue and possess physical, uncontracted, indices of dimension $2$ (represented by their dangling legs) and virtual indices of dimension $\chi$ (represented by the edges of the network) which are shared with neighboring tensors. Positive, diagonal bond tensors $\Lambda_{e}$ live on the edges $e$ between the site tensors and are coloured in grey.}
    \label{Fig:heavyhexdiagram}
\end{figure*}

The power of noise-mitigation techniques was recently demonstrated in quantum simulations of the dynamics of a two-dimensional (2D) transverse field Ising model on a $127$-qubit heavy-hexagon lattice~\cite{kim2023} (see Fig.~\ref{Fig:heavyhexdiagram}). Expectation values of a number of observables were extracted after a few Floquet cycles using zero-noise extrapolation techniques~\cite{temme17}. The experimental results were found to be much more accurate than several classical tensor network approaches applied to the same problem, even when the tensor network methods utilized significant computational resources.

In this work we adopt a tensor network approach that respects the qubit connectivity of the heavy-hex lattice to simulate the dynamics of this kicked Ising model on systems of various sizes. Our tensor network ansatz is approximately contracted with belief propagation, which works best when correlations in the system remain ``tree-like''. By quantifying these correlations we explicitly show how this assumption becomes increasingly valid as the lattice size increases. This allows our method to achieve results, for the $127$ qubit problem, to a much higher degree of accuracy than previously reported classical results or the error mitigated quantum computer. Even at large circuit depths where there are no exact results to benchmark against, we utilize extensive MPS calculations and boundary MPS approaches to provide substantial evidence that our results are still highly accurate and the correlations are tree-like. Finally, we conclude by showing how our method can be used to accurately simulate the thermodynamic limit of the problem to long-times, and therefore can simulate a high-depth quantum circuit involving an infinite number of qubits.
Our work demonstrates the effectiveness of a belief propagation tensor network approach for solving many-body dynamics problems. We anticipate our chosen methodology will find success and serve as a benchmark when applied to problems with locally tree-like correlations and limited entanglement.

\emph{Model and Ansatz.} Our focus here is on the dynamics of the Trotterized kicked transverse field Ising model given by the unitary 
\begin{align}
    &U(\theta_{h}) = \notag \\ &\left(\prod_{\langle v,v' \rangle}\exp \left({\rm i} \frac{\pi}{4}Z_{v}Z_{v'} \right)\right)\,\left(\prod_{v}\exp \left(-{\rm i} \frac{\theta_{h}}{2}X_{v}\right)\right),
    \label{Eq:Propagator}
\end{align}
where $Z$ and $X$ denote Pauli operators and $\langle v,v' \rangle$ indicates that $v$ and $v'$ are neighbors on the corresponding lattice. The lattice we are concerned with is that of the `heavy-hex' lattice which corresponds to a hexagonal lattice decorated with additional qubits along the edges (see Fig.~\ref{Fig:heavyhexdiagram}). The dynamics of this model was recently simulated on the IBM Eagle quantum processor~\cite{kim2023}, which corresponds to a lattice of $6 \times 3$ heavy-hexagons plus two additional qubits. 

In order to simulate this system on a classical computer, we adopt a tensor network approach that respects the qubit connectivity of the heavy-hex lattice  (see Fig.~\ref{Fig:heavyhexdiagram}). We fix a maximum amount of entanglement in the system by limiting the bond dimension $\chi$ of the network: each application of $U(\theta_{h})$ will double the bond dimension of the network and thus it is necessary to limit the bond dimension and perform truncations. We then evolve our tensor network state (TNS) by application of the gates in $U(\theta_{h})$ under the belief propagation (BP) approximation \cite{TensorNetworkBeliefPropagation1,TensorNetworkBeliefPropagation2,TensorNetworkBeliefPropagation3}; referring to the resulting TNS as a BP-approximated TNS. Unless otherwise stated, we also extract expectation values from the TNS using belief propagation. Explicit details of our BP-based method are provided in the \hyperref[Sec:Methods]{Methods} section and in Ref. \cite{TindallGauging2023}. 
The BP method is fully controlled on trees but incurs a potentially small but uncontrolled approximation when there are loops in the network; the error from this approximation is, in general, smaller for larger loops (see \hyperref[Sec:Methods]{Methods} for further discussion on this). 
Here, we demonstrate that for a sufficiently large lattice, even at significant circuit depths, the correlations in this model remain `tree-like' in the sense of the BP approximation giving very accurate results. Let us present these results.

\begin{figure*}[t!]
    \centering
    \includegraphics[width = \textwidth]{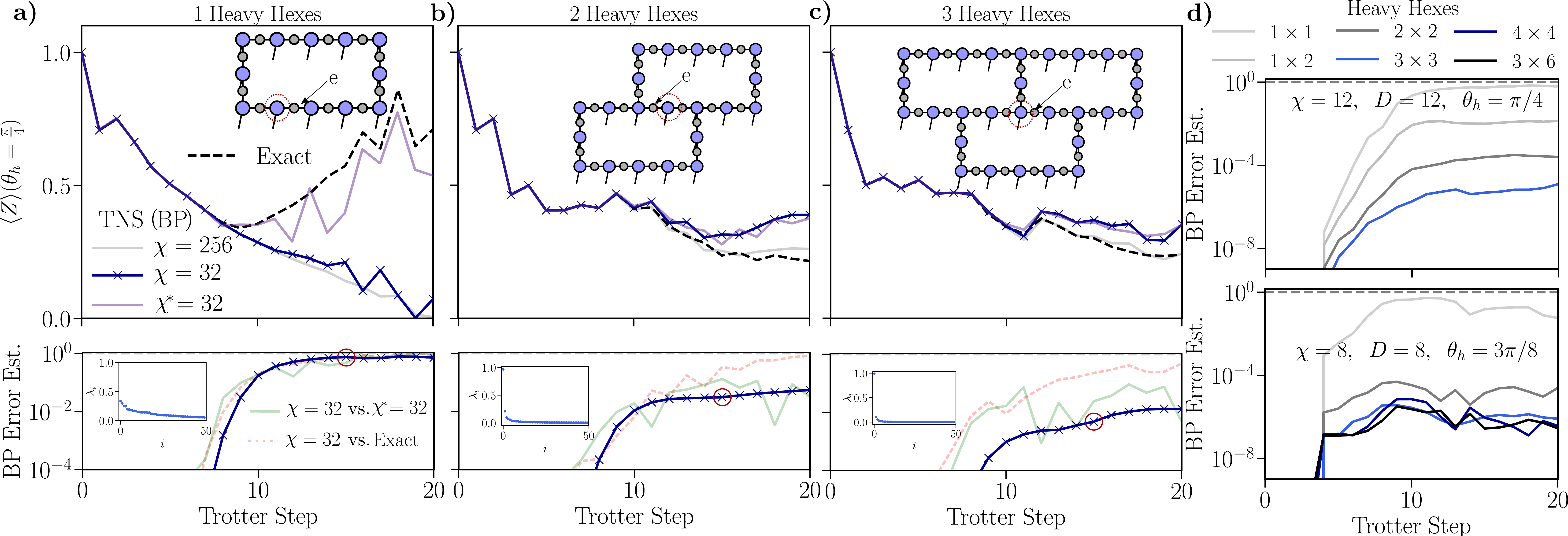}
    \caption{Dynamics of the kicked Ising model on small heavy-hex lattices of varying size. \textbf{a-c) } Results from the BP-evolved TNS for several bond dimensions and systems of one to three heavy-hexes are compared to exact state vector solutions at $\theta_{h} = \pi/4$. The results labelled with $\chi^*$ are obtained by computing expectations values of the BP-evolved TNS using exact contraction, while the results labelled with $\chi$ are obtained by computing expectations values of the BP-evolved TNS using BP contraction. In both cases, the states are evolved by applying gates using the BP approximation. Top plots show dynamics of the magnetization on the indicated (red ring) site, bottom plots show the BP error estimate (based on the spectrum of the edge environment --- see Fig. \ref{fig:EdgeEnvironment} and Eq. (\ref{Eq:BPErrorEstimate})) for the $\chi = 32$ TNS along the indicated edge $e$ versus the Trotter step. The dotted faded red line shows the relative error between the $\chi =32$ TNS magnetization approximated by BP and the exact magnetization while the faded green line shows the relative error between the $\chi =32$ TNS magnetization approximated by BP and the magnetization obtained by contracting the same TNS exactly. Insets in the bottom plots show the first $50$ singular values of the edge environment after $15$ Trotter steps. \textbf{d)} BP error estimate approximated using a boundary MPS contraction scheme (see the Appendix on \hyperref[Subsec:BoundaryMPS]{Boundary MPS} for details) of the TNS for $n \times m$ lattices of $n$ rows and $m$ columns of heavy hexes. Top) TNS with $\chi = 12$ and a boundary MPS contraction scheme with maximum MPS bond dimension $D = 12$ at $\theta_{h} = \pi/4$. Bottom) TNS with $\chi = 8$ and a boundary MPS contraction scheme with maximum MPS bond dimension $D = 8$ at $\theta_{h} = 3\pi/8$.}
    \label{Fig:smallheavyhexdynamics}
\end{figure*}

\emph{Results.} We start by considering lattices with a small number of heaxy-hexagons, where an exact state vector simulation is possible and our method can be directly benchmarked (see Fig. \ref{Fig:smallheavyhexdynamics}). Specifically, we compute the dynamics of the on-site magnetization for $\theta_{h} = \pi/2$ and lattices consisting of $1$, $2$, and $3$ heavy-hexagons respectively. We also compute the separability of the `edge environment' of the BP-approximated TNS along a chosen edge (see the \hyperref[Subsec:BPErrorEstimate]{BP error estimate} section for a definition). This environment corresponds to the contraction of the TNS down to the given edge and when it is completely separable the belief propagation assumption is exact. The separability thus gives us an estimate of the error stemming from the BP approximation.
 
For short circuit depths $n < 6$ our method gives perfect agreement with the exact simulation and the edge environments are all completely separable because the light cone of circuit does not reach around the loops. For larger circuit depths there is deviation between the exact dynamics and the  BP-approximated TNS dynamics. This can be explicitly characterized by a decrease in the separability of the edge environments, since the BP method approximates the environments used to perform the gate evolution as outer products of environments coming from incoming edges of a region of the system. Our small lattice simulations demonstrate something quite remarkable: as the number of heavy-hexagons increases, the BP approximation at fixed $\chi$ improves significantly and the edge environment becomes highly separable even out to $20$ Trotter steps. Moreover, in Fig. \ref{Fig:smallheavyhexdynamics}d) we use boundary MPS (see \hyperref[Subsec:BoundaryMPS]{Boundary MPS} for a description of the method) to show that this increased separability persists with increasing system sizes.
\par The accuracy of belief propagation does tend to improve with increasing system size (for instance, belief propagation on a finite ring directly gives the exact results of the thermodynamic limit, independent of system size) and may go some way towards explaining the increased accuracy of the BP simulation as we increase the number of heavy hexes. Nonetheless, the heavy-hex lattice still has finite loops in the thermodynamic limit and thus our results are indicative of something further: a level of interference between the loops in the lattice as we increase the system size. We note that Ising models have been observed to display some slow thermalization and confinement properties under a quantum quench \cite{Birnkammer2022, Marton2017} and we believe a similar effect is occurring here.
\par
\begin{figure*}[t!]
    \centering
    \includegraphics[width = \textwidth]{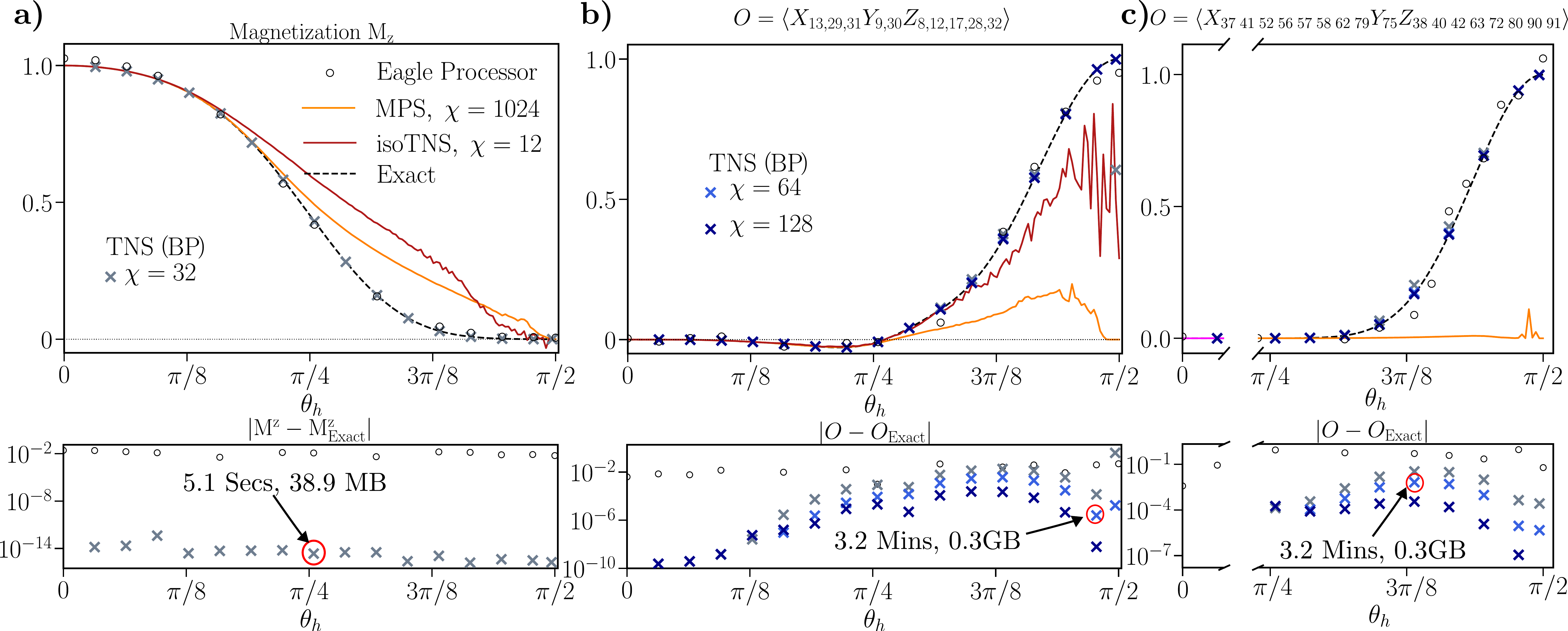}
    \caption{Comparison for classically verifiable systems of our BP-approximated tensor network state approach to simulating the dynamics of the kicked transverse field Ising model on a heavy-hex lattice versus the Eagle quantum processor and alternative tensor network methods. Expectation values with respect to the state $\vert \psi(\theta_{h}, 5) \rangle$ (i.e following $5$ Trotter steps of the dynamics of the model --- see Eqs. (\ref{Eq:Propagator}) and (\ref{Eq:State})) are plotted, alongside exact results determined from light cone simulations. \textbf{a)} Average magnetization. \textbf{b)} Weight-10 observable. \textbf{c)} Weight-17 observable. The bottom plots show errors defined as the absolute difference between the simulation result and the exact result. For some data points the error from our TNS simulation is too small to fit on the scale of the plot and so these points are not marked. Circled, annotated points denote, for a given $\theta_{h}$, the memory required to store the state of the system at the given bond dimension $\chi$ and the walltime associated with performing the simulation and calculating the relevant observable on a Macbook M1 Pro.}
    \label{Fig:comparison}
\end{figure*}

 We now consider the $127$ qubit heavy-hexagon lattice which corresponds to that of the IBM Eagle processor and we compare our method to the experimental quantum simulation results first from Fig.~3 of Ref.~\cite{kim2023}. Our results on smaller system sizes will be useful in explaining the accuray of our results here. In Fig.~\ref{Fig:comparison} we overlay the quantum simulation results with that
of our BP-approximated TNS dynamics shown as cross symbols.
Here expectation values are measured after $5$ Trotter steps and exact results, based on brute force light cone simulation techniques, are available to allow us to directly assess the errors. 
The tensor network state (TNS) and gate evolution methods we use simulate the full $127$-qubit system and result in highly accurate expectation values. In Fig.~\ref{Fig:comparison}(a) we compute the expected value of the average single-site magnetization and show that we can obtain an accuracy\footnote{In terms of absolute difference between our result and the exact value.} of $\sim 10^{-14}$ with a simulation that runs in less than $10$ seconds on a laptop computer. 
Importantly, even for Figs.~\ref{Fig:comparison}(b) and Figs.~\ref{Fig:comparison}(c), where we consider higher weight observables which could be more strongly affected by loop correlations, we are still able to calculate these observables to a remarkable accuracy using our BP-approximated TNS. Specifically,
we obtain values of these higher weight observables to orders of magnitude better accuracy than the quantum processor with a simulation that takes less than $4$ minutes (for these observables) to run on a laptop and a state that takes up, at most, $0.3$GB of memory.
The remarkable accuracy we are able to achieve corroborates with our earlier analysis of the error of the BP approximation as a function of system size (see Fig. \ref{Fig:smallheavyhexdynamics}), where we found that the BP error decreases as we increase the system size for this model and lattice.

\begin{figure*}[t!]
    \centering
    \includegraphics[width = \textwidth]{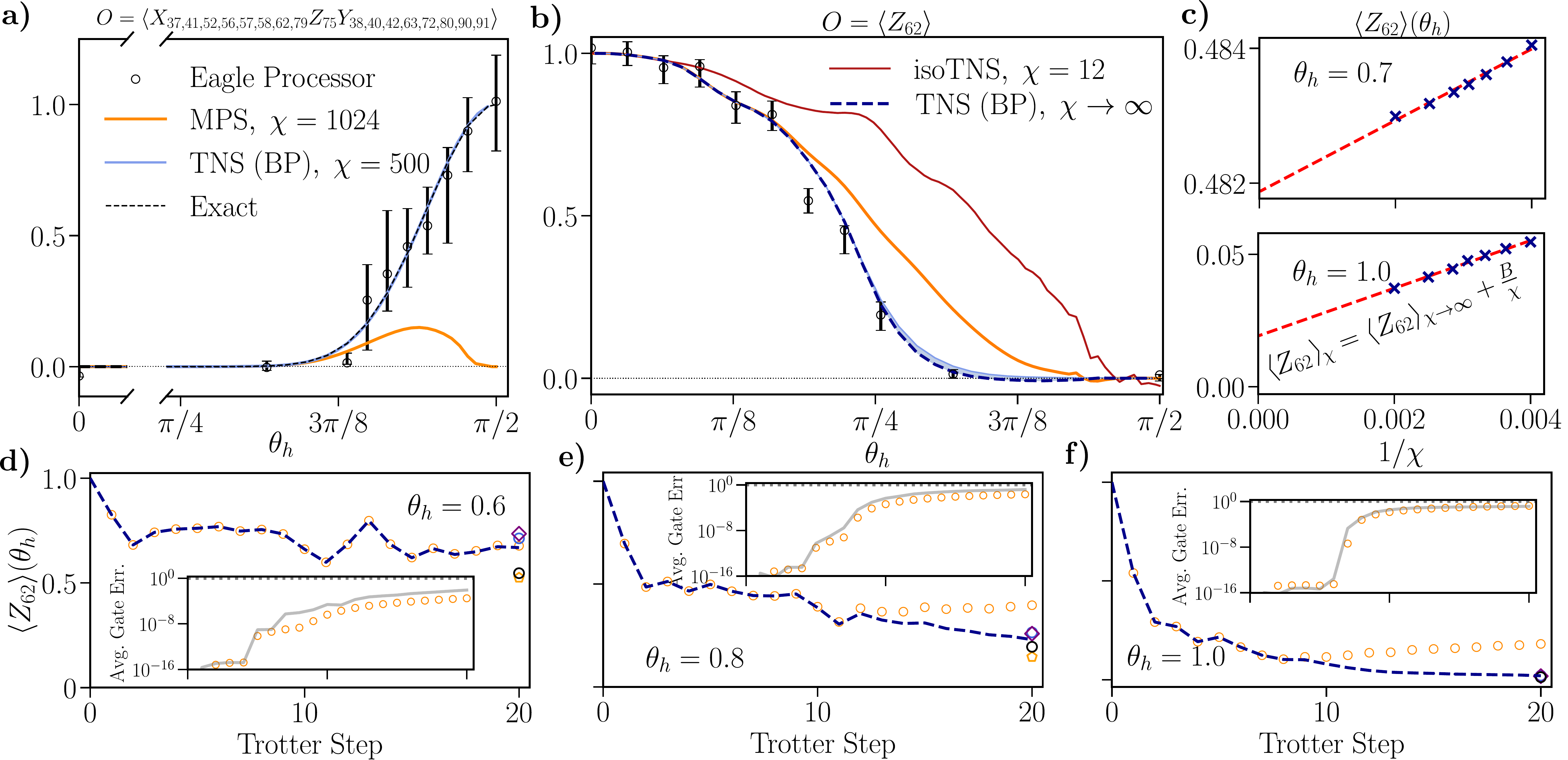}
    \caption{Comparison for deeper circuits of our BP-approximated tensor network state approach to simulating the dynamics of the kicked transverse field Ising model on the heavy-hex lattice versus the Eagle quantum processor and alternative tensor network methods. Expectation values calculated following a number of Trotter steps of the dynamics of the model --- see Eq. (\ref{Eq:Propagator}) --- are plotted. \textbf{a)} Weight-17 stabilizer after $6$ steps of evolution. Here exact data is now available \cite{Liao2023} and our BP data for $\chi = 500$ is within $10^{-4}$ of the exact result for all $\theta_{h}$. \textbf{b)} Weight-1 observable after $20$ steps. The shaded region shows the difference between our finite $\chi = 500$ bond dimension data and the data extrapolated to infinite bond dimension, where we believe the true answer lies. \textbf{c)} Top and bottom plots show observables in  \textbf{b)} at $\theta_{h} = 0.7$ and $\theta_{h} = 1.0$ respectively as a function of inverse bond dimension of the TNS. Red dashed lines represent a least squares fit of the form $A + B \slash \chi$ taken on the data, and we take $A$ to be the predicted value of the observable in the limit $\chi \rightarrow \infty$. Even in the limit $\chi \rightarrow \infty$ there will generally be some deviation from the exact result due to the BP approximation that we use for evolving the state and computing expectation values (see the \hyperref[Sec:Methods]{Methods} section). Our analysis of the errors due to BP for this system, however, suggest this deviation is likely to be very small. \textbf{d-f}) Dynamics of $\langle Z_{62} \rangle$ using the BP-approximated TNS approach versus a MPS approach (with bond dimension $2500$) with light cone depth reduction (orange) for $\theta_{h} = 0.6, 0.8$ and $1.0$ respectively. Results from other methods at depth $20$ are shown as black circles (Eagle processor \cite{kim2023}), blue circles (truncated Pauli strings \cite{Chan2023, begušić2023}), purple diamonds (TNO \cite{Liao2023}) and orange pentagons (MPO \cite{Zalatel2023}).
    Inset shows average gate error from the MPS approach (pink circles) and absolute difference between the BP-approximated TNS and the MPS result (solid grey line). }
    \label{Fig:F2}
\end{figure*}

Turning next to larger numbers of Trotter evolution steps we show results in Fig.~\ref{Fig:F2} for properties also
computed by the quantum processor. For the $n = 6$ Trotter step simulation (Fig.~\ref{Fig:F2}a), where a weight-$17$ observable is measured, exact results are now available \cite{Liao2023} and our BP-approximated TNS results at $\chi = 500$ are within $10^{-4}$ of these results for all values of $\theta_{h}$ we plot. 
For $n = 20$ (Fig.~\ref{Fig:F2}b) exact data is currently unavailable and we push to sufficiently large bond dimensions to perform an extrapolation of the results from BP-approximated TNS to infinite bond dimension; demonstrating the reliability of a linear extrapolation in $1/\chi$ for select $\theta_{h}$ in Fig.~\ref{Fig:F2}c).  We capture the known exact values at the Clifford points $\theta_{h} = 0$ and $\theta_{h} = \pi/2$. Notably, for this circuit in the region $\frac{\pi}{8} \leq \theta_{h} \leq \frac{3\pi}{8}$ there have been discrepancies between various classical methods and the quantum processor. 
In \hyperref[Subsec:Comparison]{Appendix}, we discuss the different classical simulation methods further and show a comparison in   Fig.~\ref{Fig:Z62FullComparison} of the various results \cite{kim2023, Zalatel2023, Chan2023, Liao2023, begušić2023, kechedzhi2023}. 

Beyond the results we presented above giving evidence for the separable nature of the edge environments in this system, we present further evidence that the  BP-approximated TNS method is highly accurate throughout the whole phase diagram in Fig. \ref{Fig:F2}d-f).
Specifically, we compute the dynamics at every Trotter step of $\langle Z_{62} \rangle$ for several $\theta_{h}$ and compare the BP-approximated TNS to our own MPS calculations as an independent check. Our MPS approach combines multiple non-trivial techniques:  i) utilizing light cone depth reduction to calculate $\langle Z_{62} \rangle$ at every Trotter step, ii) using a higher bond dimension than that in Ref. \cite{kim2023}, and iii) implementing an improved site ordering to lower the entanglement and gate error.
We find that the difference between the BP-approximated TNS and MPS is directly correlated with the error from the MPS method and that both methods agree closely when the error in the MPS method is itself small. 
When the MPS error is small, one can consider MPS to be exact as it makes no uncontrolled approximations.
The fact the BP-based method agrees with the MPS method when MPS exhibits very small errors suggest the BP error is also minimal. 
This is clearest for $\theta_{h} = 0.6$ in Fig.~\ref{Fig:F2}d) where we are able to push our MPS simulations to a bond dimension large enough for the average gate error to stay below $10^{-4}$. At all times for $\theta_{h} = 0.6$, there is clear agreement between BP-approximated TNS and MPS, yet disagreement at depth 20 versus the other methods shown \cite{Zalatel2023, begušić2023, Chan2023, kechedzhi2023, Liao2023}. This agreement is only possible if the state possesses tree-like correlations and thus reinforces our earlier results on the general accuracy of the BP approximation for the dynamics of this system on this lattice.
For larger $\theta_h$, in particular the $\theta_h=1.0$ results shown in Fig.~\ref{Fig:F2}f, the BP results agree with the new MPS results until about step 10 where the MPS error starts to grow. We can be confident that the discrepancy there is due to the MPS method, not BP, because MPS is a controlled method that self-reports a significant error for larger steps and because the BP agrees with the result of $\sim 0$ at step 20 which is predicted by a range of classical methods \cite{Liao2023, Zalatel2023, kechedzhi2023, begušić2023, Chan2023} due to the system being ergodic in this regime.

\begin{figure*}[t!]
    \centering
    \includegraphics[width = \textwidth]{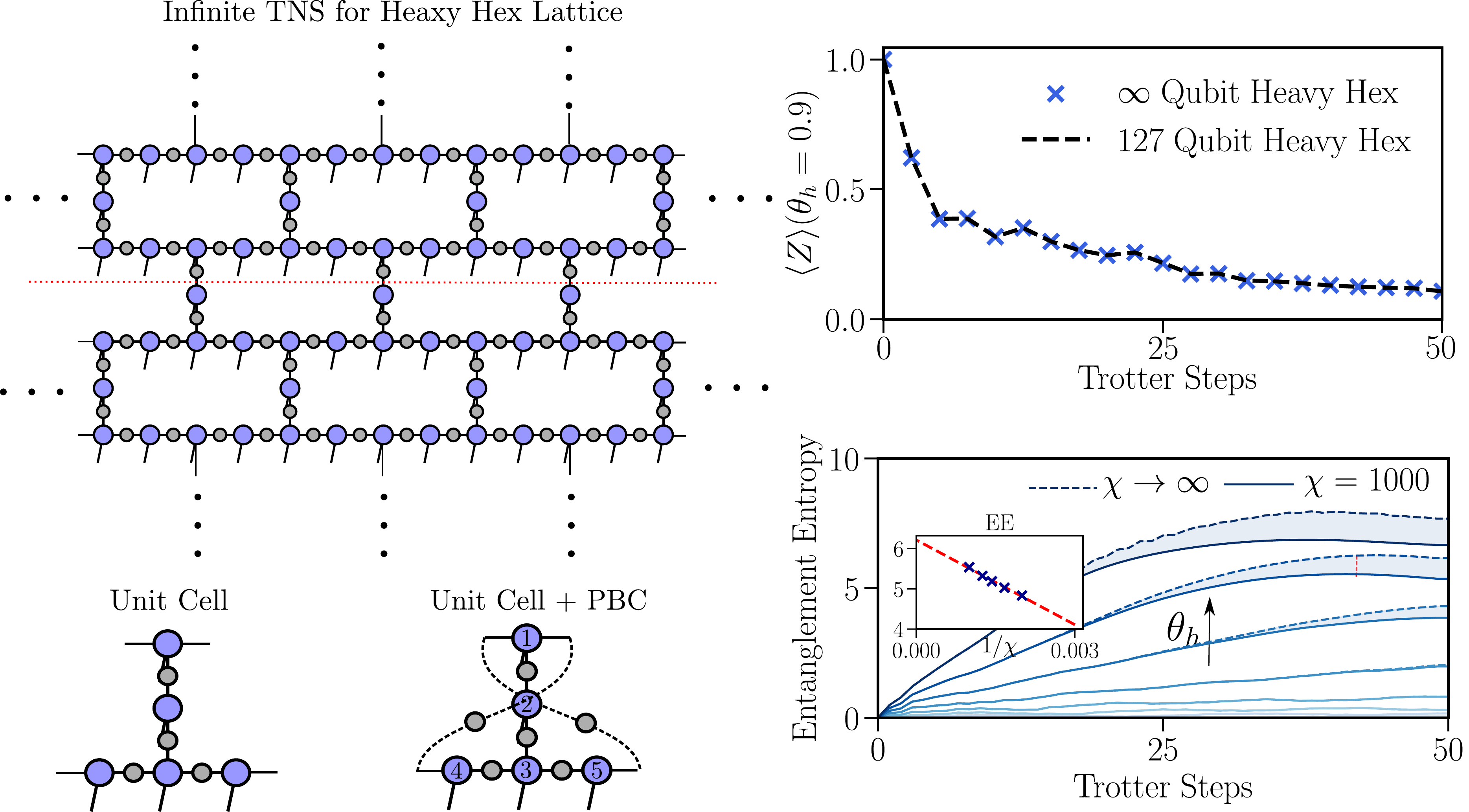}
    \caption{Tensor network state for the infinite heavy-hex lattice. The unit cell is a 5-site tensor network. By adding in appropriate periodic boundary conditions and simulating the kicked Ising model with the BP-approximated TNS method we recover results for simulation of the infinite lattice with the BP-approximated TNS method. Top Right) Dynamics of $\langle Z_{62} \rangle$ for $\theta_{h} = 0.9$ simulated using the BP-approximated TNS method at bond dimension $\chi = 400$. Crosses correspond to the dynamics of $\langle Z_{3} \rangle$  on the periodic boundary condition (PBC) unit cell (where site $3$ is marked) with $\chi = 400$ corresponding to an infinite heavy-hex lattice while the dashed line corresponds to the dynamics of $\langle Z_{62} \rangle$ on the $127$ qubit heavy-hex lattice with $\chi = 400$. Bottom) Dynamics of the bipartite entanglement entropy density $s$ (see Eq. (\ref{Eq:EntanglementEntropy}) for the definition), calculated using the infinite BP-approximated TNS method and extrapolated to infinite bond dimension. Different curves correspond to different values of $\theta_{h}$, ranging in steps of $0.1$ from $0.1$ to $0.9$. The partition is shown by the red dotted line on the infinite lattice. The shaded area shows the difference between the extrapolated result (dotted line) and the finite $\chi = 800$ result. The inset shows an example extrapolation in $1/\chi$ for $\theta_{h} = 0.8$ after $40$ Trotter steps.
    }
    \label{Fig:F4}
\end{figure*}

\emph{Dynamics of the infinite heavy-hex lattice}  - One of the powerful features of tensor network methods is their ability simulate \emph{infinite} lattice models as long as they possess some form of translational invariance \cite{Vlaar2021, Orus2009, Phien2015a}. Here we present results on the dynamics of the kicked transverse field Ising model on an infinite heavy-hexagon lattice, corresponding to a quantum computer with an infinite number of qubits. Again, we approximate the dynamics and take expectation values using the BP approximation. Given the evidence that we have presented on the accuracy of the BP approximation for the finite case of this system, especially for larger system sizes, we expect that these results are highly accurate.

For the infinite heavy-hex lattice there is a $5$-site unit cell which can be tiled to produce the infinite lattice (see Fig. \ref{Fig:F4}). 
It can be shown that simulating a periodic system on the unit cell using belief propagation corresponds to simulating the infinite lattice under the belief propagation approximation, where the sites of the periodic system constitute a unit cell of the infinite system \cite{TindallGauging2023}.
This is analogous to the standard approach to simulating infinite systems with the simple update tensor network method \cite{Jiang2008}. 
Therefore we take the single unit cell, impose periodic boundary conditions, and present results for the belief propagation approximated dynamics.
Fig. \ref{Fig:F4} illustrates this idea, showing the dynamics of the magnetization of the infinite system compared to the expected magnetization of a representative site near the center of the finite system. The extremely close agreement between the magnetization of the infinite and finite heavy-hex lattices provides strong evidence that boundary effects are minimal in the $127$-qubit model and the results are already very close to those of the thermodynamic limit. 

In Fig. \ref{Fig:F4} we also show the time-dependence of the bipartite entanglement entropy per edge $s$ across a bipartition of the infinite lattice. Our infinite BP-approximated TNS method gives us an estimate of this quantity via the spectrum $\{\lambda_{i}\}$ of the bond tensors along the edges separating the partitions:
\begin{equation}
    s = \sum_{i=1}^{\chi^{2}}\lambda_{i}^{2}{\rm log_{2}}(\lambda_{i}^{2})
    \label{Eq:EntanglementEntropy}
\end{equation}
which requires summing over the spectrum of only one of the bond tensors due to the fact that they are all identical along the cut being made. We observe that the entanglement in the system shows a sharp linear growth at short times before slowing down significantly and potentially saturating over a large time scale. This long-lived plateu is consistent with studies of the decay of the magnetisation in a smaller system \cite{kechedzhi2023} which suggests that the time-to-decay scales exponentially with the inverse of $\theta_{h}$.
This long-time slow growth means that we can accurately simulate the infinite quantum processor up to large circuit depths for smaller values of $\theta_{h} \lesssim \pi/4$. For larger values of $\theta_{h}$, the entanglement grows sufficiently quickly that, with our current resources, we are unable to accurately determine the entanglement entropy in the system beyond $\sim 25$ Trotter steps.

\emph{State ansatz.} Our ansatz for the wave function of the system is a tensor network which directly reflects the `heavy-hex' qubit connectivity of the processor (see Fig.~\ref{Fig:heavyhexdiagram}). The physical properties of a tensor network are invariant under a gauge symmetry corresponding to the insertion of any invertible matrix $G_{e}$ and its inverse $G_{e}^{-1}$ on any contracted, internal bond/edge $e$ of the network.
We take advantage of this symmetry to keep the tensor network in the \emph{Vidal gauge} \cite{Vidal2003, Jiang2008}. This gauge corresponds to the choice of positive, diagonal bond tensors $\Lambda_{e}$ residing on the edges of the network and the on-site tensors $\Gamma_{v}$ of the network obeying certain isometric properties (see Eqs. (\ref{Eq:IsometryPropertyDefinition}) and (\ref{Eq:IsometryDefinition})). These isometric properties are important for maintaining accuracy during the evolution of the network. We use $\vert \psi(\theta_{h}, n) \rangle$ to denote the TNS of the system after $n \geq 1$ applications of $U(\theta_{h})$, i.e. 
\begin{equation}
    \vert \psi(\theta_{h}, n) \rangle = \left(\prod_{i=1}^{n}U(\theta_{h})\right)\vert \psi(0) \rangle = U^{n}(\theta_{h})\vert \psi(0) \rangle,
    \label{Eq:State}
\end{equation}
where $\vert \psi(0) \rangle$ is the initial state of the system. We use the same initial state as in Ref. \cite{kim2023}: $\vert \psi(0) \rangle = \vert\! \uparrow \uparrow \hdots \uparrow \rangle$.
The single-site $X$ rotations in $U(\theta_{h})$ can be applied to $\vert \psi(\theta_{h}, n) \rangle$ exactly and the two-site gates are applied approximately using the simple update \cite{Jiang2008} procedure (see the \hyperref[Sec:Methods]{Methods} section), which involves truncating the internal indices of the TNS to keep them less than or equal to a prescribed maximum bond dimension $\chi$. 
Following a single Trotter step, the tensor network is regauged using belief propagation, a well established statistical inference algorithm which can  be formulated for tensor networks \cite{TensorNetworkBeliefPropagation2} which we find improves the accuracy of the simple update procedure \cite{Jahromi2019, TindallGauging2023}. Regauging before applying each gate with simple update would be most accurate and would be equivalent to performing each gate application with environments computed from the BP fixed point, however in practice we find that is too computationally expensive. Regauging after each Trotter step provides a good balance between accuracy and speed. Belief propagation allows us to rapidly determine the necessary transformation matrices by performing \emph{message passing} over the network representing $\langle \psi(\theta_{h}, n) \vert \psi(\theta_{h}, n) \rangle$. The use of belief propagation as a method to efficiently `regauge' a tensor network was recently formulated by some of the current authors in Ref. \cite{TindallGauging2023} and is closely related to other known gauging methods  \cite{Ran2012, Phien2015, Jahromi2019, SimpleUpdateGauging3}. We emphasize that the results presented here could have been achieved with those known gauging methods.

\emph{Measuring expectation values.} We measure single-site expectation values of the gauged state $\vert \psi(\theta_{h}, n) \rangle$ using a rank-one approximation for the environments of local regions of the network (see \hyperref[Sec:Methods]{Methods} for more details). Such a method is only guaranteed to be exact in the limit of a tree network. Nonetheless, as demonstrated by our results here, the large loop structure of the 
heavy-hexagon lattice and dynamics of the model means that the network is locally tree-like and therefore amenable to such an approximation. In Ref.~\cite{kim2023} specific higher-weight observables were also measured to highlight that non-trivial results can be obtained in the ``regime of strong entanglement'' --- such observables can be difficult to accurately measure for loopy tensor networks. 
We can, however, exploit the Clifford properties of the circuit at $\theta_{h} = \frac{\pi}{2}$ to transform the problem of measuring higher-weight observables after $n$ Trotter steps into one of measuring a single-site observable after evolution by $n$ Trotter steps with the propagator $U(\theta_{h})$ and $n$ Trotter steps with the propagator $\left(U(\frac{\pi}{2})\right)^{\dagger}$. 
Specifically, the operator $U^{n}\left(\frac{\pi}{2}\right)Z_{i}\left(U^{n}\left(\frac{\pi}{2}\right)\right)^{\dagger}$ is always a single Pauli string and its expectation with respect to $\vert \psi(\theta_{h}, n) \rangle$ can be obtained by further evolving $\vert \psi(\theta_{h}, n) \rangle$ with $\left(U(\frac{\pi}{2})\right)^{\dagger}$ and then measuring $Z_{i}$. For instance,
\begin{equation}
    X_{13, 29, 31}Y_{9,30}Z_{8,12,17,28,32} =  U^{5}\left(\frac{\pi}{2}\right)Z_{13}\left(U^{5}\left(\frac{\pi}{2}\right)\right)^{\dagger},
\end{equation} 
meaning this observable can be calculated, with respect to $\vert \psi(\theta_h,5) \rangle$, by measuring $\langle Z_{13} \rangle$ with respect to the state $\left( U^{5}\left(\frac{\pi}{2}\right)\right)^{\dagger}\vert \psi(\theta_h,5) \rangle$. We can reach this state straightforwardly by performing a $10$ ($5$ + $5$) Trotter step tensor network simulation. We use this `extended time evolution' method to measure all higher-weight observables presented in Ref. \cite{kim2023} and note that this procedure is entirely generic: any long Pauli string can be generated out of a single Pauli by application of a Clifford circuit. Naturally this extended time evolution further increases the entanglement we must deal with as each application of $U\left(\frac{\pi}{2}\right)$ will double the bond dimension if no truncation is performed. Thus an upper bound on the bond dimension $\chi_{\rm max}$ needed to evolve the system by $n$ Trotter steps and the measure a string which can be generated by $n'$ applications of $U\left(\frac{\pi}{2}\right)$ to a single-site Pauli operator is $2^{n + n'}$. Generically, longer strings will require larger values of $n'$. Figures \ref{Fig:comparison} and \ref{Fig:F2} here, however, show that we can use values of $\chi \ll \chi_{\rm max}$ and still get very accurate results.

We would also like to emphasize that these higher-weight observables could be measured with other tensor network methods. For instance, the planar nature of the TNS on the heavy-hex lattice means the boundary MPS \cite{Verstraete2004} method can be directly applied to the norm of the state $|\psi(\theta_h,n) \rangle$ to measure a desired higher-weight observable. We used the boundary MPS method in this paper to approximately compute the edge environment and also found extremely close agreement between BP and boundary MPS when measuring single-site observables (see \hyperref[Subsec:BoundaryMPS]{Boundary MPS}).

\emph{Conclusion.} 
We have shown that a 127 qubit simulation of Floquet dynamics of the kicked Ising model on a heavy-hexagon lattice, recently simulated on a quantum processor in Ref.~\cite{kim2023}, can be performed accurately and with minimal computational resources with tensor networks. Our work  demonstrates the importance of adopting a tensor network ansatz which  reflects the spatial connectivity and entanglement structure of the system. The chosen gate evolution method for this ansatz is based on contracting the network under the belief propagation approximation. The computational scaling is $\mathcal{O}(L \chi^{4})$ for a given Trotter step evolution, where $L$ is the number of qubits and $\chi$ is the bond dimension. 
Like the commonly used and closely related simple update gate evolution method, this method directly assumes the lattice is locally tree-like, in other words assuming that loops have a minimal effect on the local properties of the network. We have presented evidence that this assumption becomes increasingly valid with increasing system size. This makes our gate evolution method highly accurate and reveals a striking loop-free behaviour to the dynamics of this kicked Ising model. We leveraged this understanding of the model and our method to perform accurate simulations of the long time dynamics of the model on an infinitely large heavy-hex lattice, corresponding to a quantum processor with an infinite number of qubits. Following an initial linear growth in entanglement, we find a remarkable result that the entanglement of the system appears to saturate over a large timescale. 
\par Looking forward, we expect that the belief-propagation based method employed here will allow efficient, accurate tensor network simulations of a range of dynamics problems and not just the one considered here. There are a number of geometries---especially in dimensions higher than two where more neighbors typically causes more mean-field-like behavior \cite{Vlaar2021}---where the effect of loops is small and we anticipate our method will find success. Moreover, while it is typically difficult to quantify the error stemming from the BP approximation, here we have performed a number of supplemental calculations (i.e. state vector calculations for small lattices, matrix product state simulations, and quantification of the BP error for increasing system sizes) to ascertain the accuracy of our results and we believe that such a methodology will be fruitful when considering other systems with a BP-based method. 
\par Another takeaway from our results is that there can be many complementary routes toward classical simulation of quantum many-body systems, especially for those with physical structure such as separation of energy scales or locality. In addition to well-known properties such as low entanglement, shallow circuit depth, or low T-gate count (in the case of nearly Clifford circuits), our work highlights the importance of 
lattice topology 
as another key property that can be exploited. We also emphasize that tensor network methods are not limited to one- and two-dimensional systems, and higher-dimensional, non-planar systems can actually become more mean-field-like, allowing tensor network approaches to again work well~\cite{tindall2022quantum, searle2023exact}. 

Finally, we would like to comment that our work opens up new directions in which highly flexible and computationally inexpensive tensor network approaches can be used to benchmark new quantum processor designs and can better delineate which many-body quantum systems could become difficult for classical computing techniques. The software we used is part of the forthcoming ITensorNetworks.jl package \cite{ITensorNetworks} being developed at the Flatiron Institute  (see the \hyperref[Subsec:ComputingResources]{Computing resources and software packages} section), which enables rapid testing and deployment of tensor network methods on arbitrary graphs. This software is continually being developed with the aim of tackling the type of simulation presented here.

\section*{Methods}
\label{Sec:Methods}
\subsection*{The Vidal gauge}
Our tensor network ansatz for the state of the system consists of both local tensors $\Gamma_{v}$ on the sites of the network and bond tensors $\Lambda_{e}$ which live on the edges of the network. The state is in the `Vidal' gauge which is characterized by a set of isometric constraints on certain groupings of local tensors. Specifically, for every edge $e = (v, v_{i})$ connecting vertices $v$ and a neighboring vertex $v_i$, if we group the following tensors
\begin{equation}
    \adjincludegraphics[valign=c]{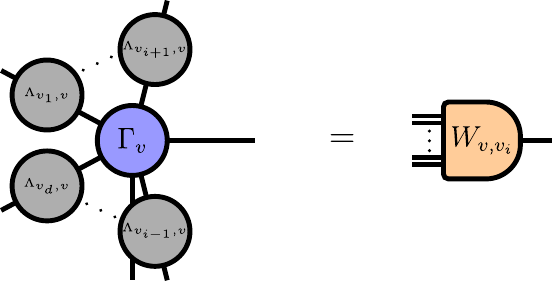} \quad,
    \label{Eq:IsometryPropertyDefinition}
\end{equation}
they form an isometric tensor obeying
\begin{equation}
    \adjincludegraphics[valign=c]{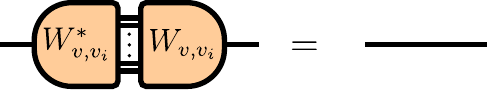} \quad ,
    \label{Eq:IsometryDefinition}
\end{equation}
where the right-hand side denotes the identity matrix and we have used the set $\{v_{1}, v_{2}, \hdots v_{d}\}$ to denote the $d$ neighbors of vertex $v$. Having the TNS obey this property is important for maintaining accuracy when applying two-site gates and when taking expectation values. Below we detail the methods we use for performing these operations and then describe how we use belief propagation to maintain the Vidal gauge during our simulations. We emphasize that working in the Vidal gauge isn't strictly necessary, and one can perform all of the same operations (gate application, computing expectation values, etc.) and get the same results without transforming to the Vidal gauge by just using message tensors found from performing belief propagation on the TNS in an \textit{arbitrary gauge} \cite{TindallGauging2023}. This is why we opt to name our tensor network state as a `BP-approximated TNS' and will discuss further in the sections below. 

\subsection*{The simple update procedure}
\begin{figure}[t!]
    \centering
    \includegraphics[width =\columnwidth]{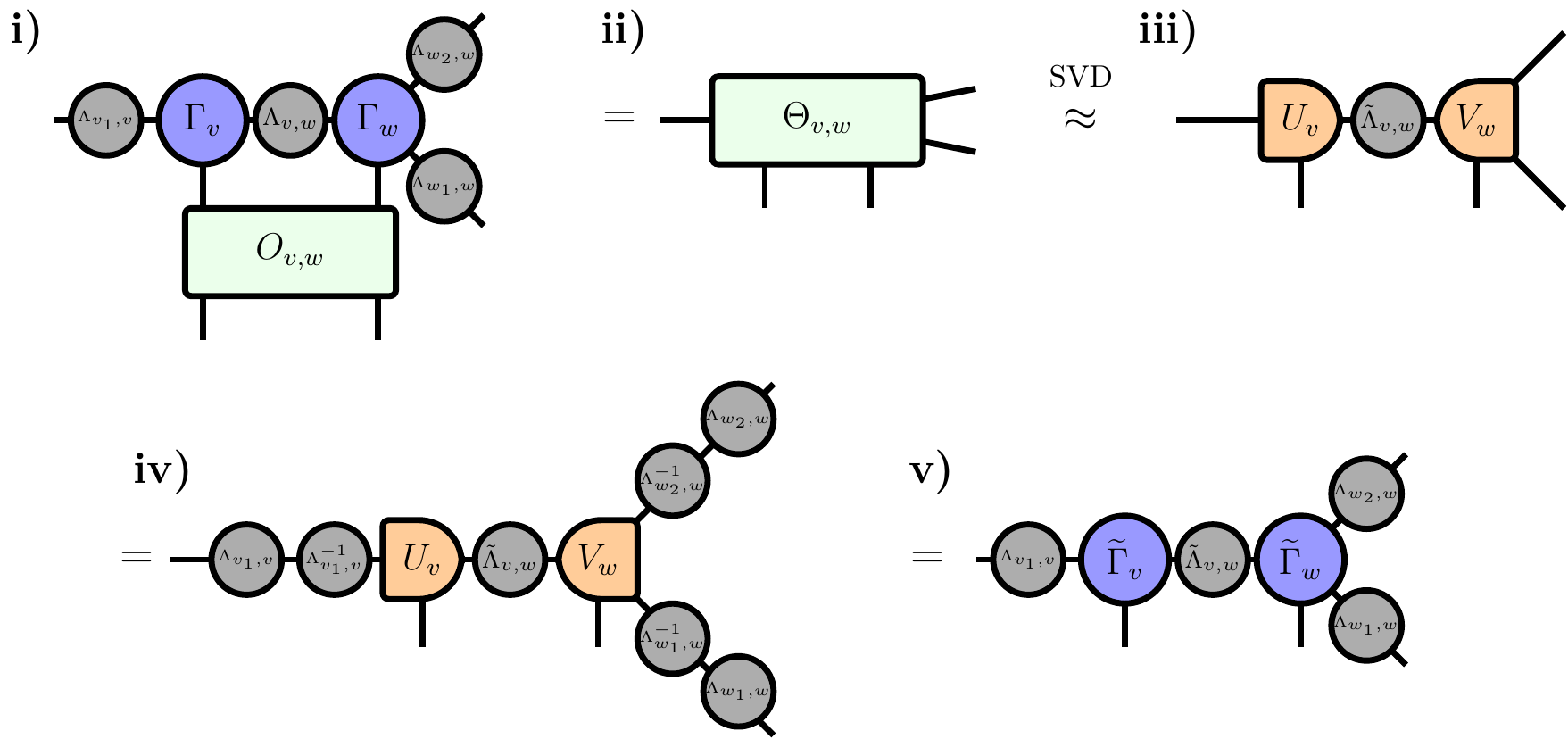}
    \caption{Steps to perform a `simple update' on the edge of a TNS in the Vidal gauge. The example pictured is two neighboring sites of degree $2$ (site tensor $\Gamma_v$) and $3$ (site tensor $\Gamma_w$) respectively. \textbf{i) - ii)} The bond tensors, site tensors, and gate are combined into a single composite tensor $\Theta_{v,w}$. \textbf{ii) - iii)} A SVD is performed on $\Theta_{v,w}$ and singular values of the resulting bond tensor $\tilde{\Lambda}_{v,w}$ can be discarded in order to truncate the bond. \textbf{iii) - iv)} Resolutions of identity, using the original bond tensors, are inserted on the exposed edges. \textbf{iv) - v)} The inverse bond tensors are absorbed, resulting in the updated local tensors $\tilde{\Gamma}_{v}$ and $\tilde{\Gamma}_{w}$.}
    \label{fig:SimpleUpdate}
\end{figure}

To apply two-site gates to the tensor network state we adopt the simple update procedure \cite{Jiang2008}. The procedure is depicted diagramatically in Fig. \ref{fig:SimpleUpdate}. For better efficiency, we use the `reduced tensor' variant of simple update \cite{Wang2011,Li2012} (not shown in Fig. \ref{fig:SimpleUpdate}). The simple update procedure can be performed on a TNS in an arbitrary gauge by working with the fixed point message tensors found from belief propagation on the TNS and updating the message tensor on the edge where the gate is applied with the bond matrix returned from the SVD procedure \cite{TindallGauging2023}. This is similar to previous work in Ref. \cite{TensorNetworkBeliefPropagation2} where message tensors were used for energy optimization.

\subsection*{Measuring expectation values within the Vidal gauge}
In order to measure a single-site observable $\langle O_{v} \rangle$ on site $v$ of our TNS we absorb the neighboring bond tensors onto the on-site tensor $\Gamma_{v}$ and contract the result with its conjugate, inserting the single-site observable $O_{v}$ along the physical index which is being contracted over.
For the example of a site with $3$ neighbors this can be visualized as
\begin{equation}
    \adjincludegraphics[valign=c]{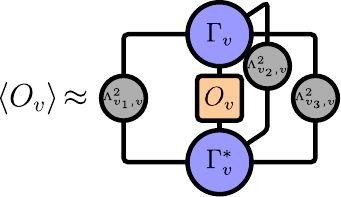} \quad,
    \label{Eq:SingleSIte}
\end{equation}
where we have defined the square of the bond tensors
\begin{equation}
    \adjincludegraphics[valign=c]{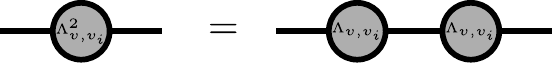} \quad.
    \label{Eq:SquareBondTensors}
\end{equation}
Equation (\ref{Eq:SingleSIte}) approximates $\langle O_{v} \rangle$ by treating the environment as a tensor product of environments coming from each of the neighbors of $v$. If the tensors of the network obey the Vidal gauge conditions (Eqs. \ref{Eq:IsometryPropertyDefinition} and \ref{Eq:IsometryDefinition}), this approximation is equivalent to computing the expectation value in an arbitrary gauge by using the fixed point message tensors of belief propagation as approximations of the environments \cite{Alkabetz2021, TindallGauging2023}. For networks that are locally tree-like, this can provide very good approximations for local observables. We find evidence that this approximation holds very well for the model and lattice studied in this work.
In the section below we detail how we use belief propagation to maintain the Vidal gauge during our simulations. This helps maintain accuracy in the simple update procedure and when taking expectation values.

\subsection*{Belief propagation on a tensor network state}
At the heart of our method for maintaining the gauge properties of our tensor network state is belief propagation (BP). BP is a well-established technique for approximating the marginals of the probability distributions of graphical models \cite{BeliefPropagation1} and has recently gained interest in the context of contracting tensor networks \cite{TensorNetworkBeliefPropagation1,TensorNetworkBeliefPropagation2,TensorNetworkBeliefPropagation3,SimpleUpdateGauging3}. 

To perform belief propagation on a TNS $\vert \psi \rangle$ in the Vidal gauge we first absorb the square roots of the bond tensors onto the $\Gamma_{v}$ tensors. For instance, for the example of a tensor with $3$ neighbors we define
\begin{equation}
    \adjincludegraphics[valign=c]{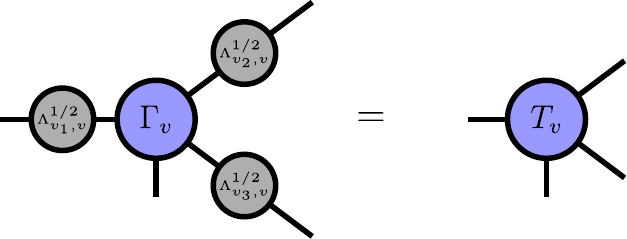} \quad.
    \label{Eq:SymmetricTensor}
\end{equation}
We then form the closed network which respresents the square norm $\langle \psi \vert \psi \rangle$ of $\vert \psi \rangle$. This `norm' network consists of the on-site tensors $\mathcal{T}_{v} = \sum_{s_v} (T^{s_v}_{v})(T^{s_v}_{v})^*$, where the summation is over the external indices $s_{v}$ of the tensors $T_{v}$ and $\left( T_{v} \right)^{*}$ (though in practice we keep the tensors separate for efficiency). 

We next define a series of `message tensors' over the the norm network, with $M_{v, v_{i}}$ denoting the message tensor directed along the edge from $\mathcal{T}_{v}$ to its neighbor $\mathcal{T}_{v_{i}}$. The indices of $M_{v, v_{i}}$ match the indices shared by tensors $\mathcal{T}_{v}$ and $\mathcal{T}_{v_{i}}$. Here, the direction of the edge is important and, generally, $M_{v,v_{i}} \neq  M_{v_{i},v}$.
A set of self-consistent equations is defined for the message tensors:
\begin{equation}
    M_{v, v_{i}} = \left(\prod_{j \in \{1, 2, ..., d \} \setminus \{i \}}M_{v_{j}, v}\right)\mathcal{T}_{v}
    \label{Eq:MessageTensorSelfConsistency}
\end{equation}
where the product runs over all $d$ neighbors of $v$ excluding $v_{i}$ and multiplication of two tensors implies a contraction over any common indices they share.
This equation can be expressed diagrammatically as
\begin{equation}
    \adjincludegraphics[valign=c]{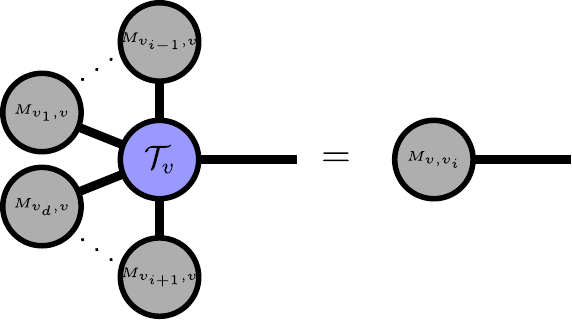} \quad.
    \label{Eq:BPEquation}
\end{equation}
Initializing the message tensors, one can iterate these equations in an attempt to converge them. The converged messages then form a rank-one approximation of the exact environment for a given tensor $\mathcal{T}_{v}$, and can be used to approximate expectation values of the TNS. The procedure for taking an expectation value in Eq. (\ref{Eq:SingleSIte}) using a state in the Vidal gauge is equivalent to taking the same state in an arbitrary gauge and approximating the environment surrounding the site with the BP message tensors \cite{Alkabetz2021, TindallGauging2023}. The more tree-like the network, the better the approximation. 

Importantly, the BP message tensors can also be used to directly define a gauge transformation which, when applied to a TNS, brings it into the Vidal gauge and guarantees the satisfaction of Eqs. (\ref{Eq:IsometryPropertyDefinition}) and (\ref{Eq:IsometryDefinition}) along all edges of the network. 
A detailed description of this method, including extensive benchmarking and discussion of its relation to other gauging methods, is given in Ref. \cite{TindallGauging2023}. 

In our simulations here we perform belief propagation gauging after every Trotter step. This is to maintain accuracy in these procedures. We should note that for shorter depth circuits, such as those simulated in Fig. \ref{Fig:comparison}), gauging after every Trotter step does not have a significant affect on accuracy. The circuit here is not deep enough to significantly alter the isometric condition in Eq. (\ref{Eq:IsometryDefinition}) and affect the simple update procedure. For longer depth circuits, such as the one run in Fig. \ref{Fig:F2}b), gauging every Trotter step is important and not performing gauging during the simulation leads to a significant loss in accuracy. 

We also always gauge the state before taking expectation values (which, as we have discussed, is equivalent to computing the expectation values with fixed point BP message tensors). Even for the shorter circuits in Fig. \ref{Fig:comparison}, not performing gauging prior to taking an expectation value can noticeably affect the accuracy of the result --- unless the bond dimension used is large enough for the simulation to be exact and so the simple update procedure preserves the gauge properties of the TNS.

\subsection*{Estimating the error of BP for general tensor networks}
\label{Subsec:BPErrorEstimate}
\begin{figure*}[t!]
    \centering
    \includegraphics[width =\textwidth]{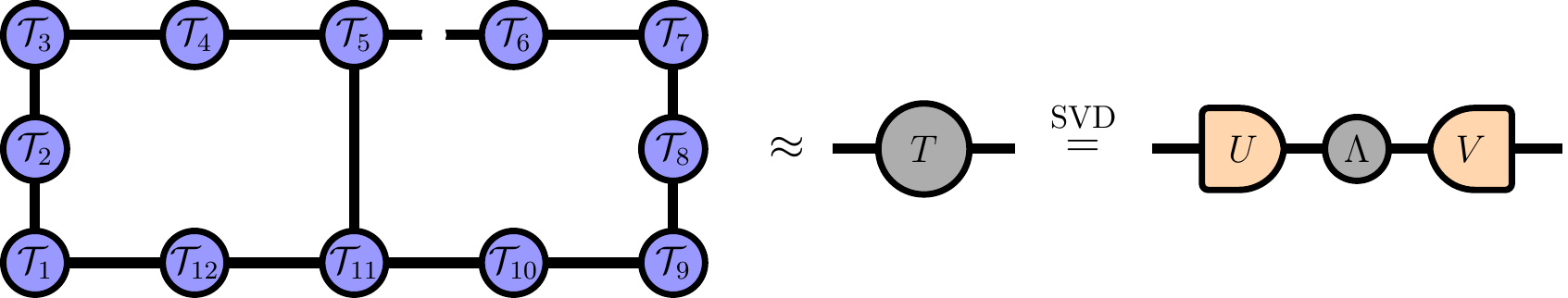}
    \caption{Forming the edge environment from the norm network of a tensor network state. One of the edges $e$ is split and all other indices are network are contracted over, reducing the cut network to a single matrix where a singular value decomposition can be performed.}
    \label{fig:EdgeEnvironment}
\end{figure*}

In order to quantify the error that belief propagation makes when approximating contractions of the network we compute the $\chi^{2} \times \chi^{2}$ `edge environment' associated with cutting the norm network along a given edge. The edge environment has previously been used in other contexts, such as improving the performance of periodic MPS methods \cite{Pippan2010} and fixing the gauge and performing truncations of general tensor networks \cite{Evenbly2018}. Here we use it to define a measure of the error of belief propagation. Specifically, we choose an edge of the norm network of our tensor network state and split the corresponding index (which is formed from the product of the bra and ket indices of that edge) --- contracting the network down to a single matrix. This process is pictured in Fig. \ref{fig:EdgeEnvironment} for an example network.

Belief propagation essentially corresponds to assuming this edge environment is rank-1 along every edge of the network, i.e. that the spectrum of the positive diagonal matrix of singular values $\Lambda$ is $\{\lambda_{1}, \lambda_{2}, ..., \lambda_{\chi^{2}}\} = \{1, 0, ..., 0\}$. We can thus estimate the error from belief propagation along an edge $e$ as
\begin{equation}
    1 - \sqrt{\frac{\lambda^{2}_{1}}{\sum_{i=1}^{\chi^{2}}\lambda^{2}_{i}}},
    \label{Eq:BPErrorEstimate}
\end{equation}
where the quantity inside the square root is known as the `index of separability' and is an established measure of the separability of a matrix based on a singular value decomposition \cite{Shihab2001}. Although the singular values are dependent on the gauge of the tensor network, we fix to the symmetric gauge \cite{TindallGauging2023} when calculating them for consistency.
This is the quantity we compute in Fig. \ref{Fig:smallheavyhexdynamics} for the pictured networks and given edges. We also include examples of spectra which correspond to a variety of separabilities in the insets. The BP error measure defined in Eq. (\ref{Eq:BPErrorEstimate}) is in the same spirit as the `cycle entropy' defined in Ref. \cite{Evenbly2018}. In general for this model we find the choice of edge is unimportant and does not qualitatively change the results we observe.
\par It is worth making a further comment on the BP error in Eq. (\ref{Eq:BPErrorEstimate}). In general, computing it is very costly due to the need to contract the full norm network, even if it is contracted approximately. One can observe, however, that generically the error will decrease exponentially in the smallest loop size of the lattice. This is most straightforwardly seen by considering a norm network which is a single translationally invariant periodic ring (loop) of size $l$. Assuming the spectrum of the on-site transfer matrix $\mathcal{T}$ is gapped then the BP error of the ring will scale as $\mathcal{O}\left(\exp(-cl)\right)$ for some constant $c$ related to the correlation length. This scaling should be generic for a lattice with smallest loop size $l$ and gapped loop correlations.

\subsection*{Computing resources and software packages}
\label{Subsec:ComputingResources}
The code used to produce the numerical results in this paper was written using the \textbf{ITensorNetworks.jl} package \cite{ITensorNetworks} --- a general purpose and publicly available Julia \cite{bezanson2017julia} package for manipulating (gauging, contracting, partitioning, evolving, etc.) tensor network states of arbitrary geometry. It is built on top of \textbf{ITensors.jl} \cite{itensor-r0.3}, which provides the basic tensor operations. 
Code is available in the current version of ITensorNetworks.jl for performing belief propagation, gauging, and the simple update procedure on arbitrary tensor network states. An example script is also included for specifically simulating the model in this paper with our BP-approximated TNS approach.
All data was produced using a single node of Flatiron Institute's Rusty computing cluster. Timings and memory usage quoted for Fig. \ref{Fig:comparison}, however, are based on the same code being run on a Macbook M1 pro. The tensor network diagrams in this paper were produced using the publicly available package \textbf{GraphTikz.jl} \cite{GraphTikz}, a general-purpose Julia package for visualizing graphs, including tensor networks. Data for the $127$-qubit simulations is currently available at: \href{https://github.com/JoeyT1994/BP-TNS-Data}{https://github.com/JoeyT1994/BP-TNS-Data}.

\section*{Acknowledgements}
J.T., M.F., M.S., and D.S. are grateful for ongoing support through the Flatiron Institute, a division of the Simons Foundation. D.S. was supported by AFOSR: Grant FA9550-21-1-0236. We would like to acknowledge Michael Zalatel, Sajant Anand, Garnet Chan, Ariel Stolbun, Andy Millis, Antoine Georges, and Bo Xiao for insightful discussions and helpful comments on the draft. We would also like to thank Michael Zalatel, Sajant Anand, Garnet Chan, Tomislav Begušić, and Haijun Liao for sharing their data with us. 

\renewcommand{\theequation}{A\arabic{equation}}
\setcounter{equation}{0} 

\section{Appendix A: Comparison of methods for calculating $Z_{62}$}
\label{Subsec:Comparison}

\begin{figure*}[t!]
    \centering
    \includegraphics[width = \textwidth]{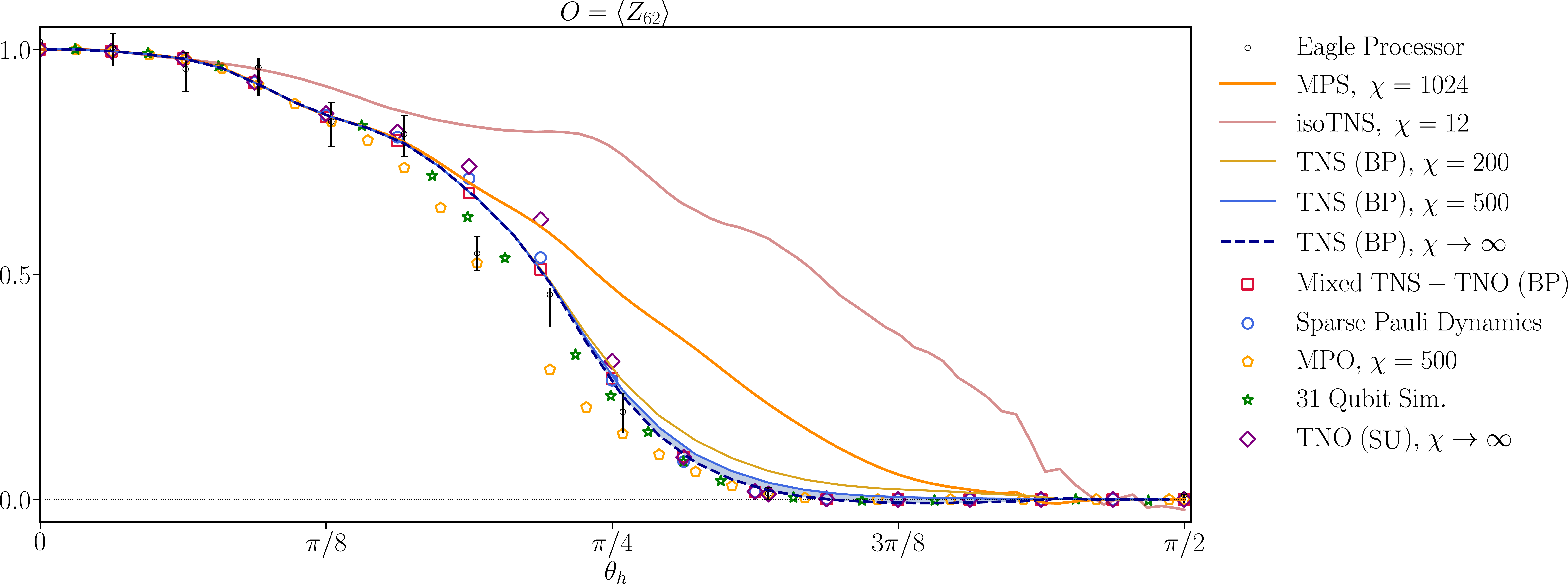}
    \caption{Comparison of various approaches to simulating the kicked transverse field Ising model on the $127$-qubit Eagle processor geometry (Fig. \ref{Fig:heavyhexdiagram}). Approaches include a quantum processor (\textbf{Eagle Processor} \cite{kim2023}), \textbf{MPS} \cite{kim2023} and \textbf{isoTNS} \cite{kim2023} methods using Schr\"{o}dinger evolution, a \textbf{TNS (BP)} approach using Schr\"{o}dinger evolution and evolved with BP (this work), a \textbf{Mixed TNS-TNO (BP)} \cite{begušić2023} approach using a combination of Schr\"{o}dinger and Heisenberg evolution and evolved with BP, \textbf{Sparse Pauli Dynamics} \cite{Chan2023, begušić2023}, a \textbf{MPO} \cite{Zalatel2023} approach using Heisenberg evolution, a $31$-qubit full state simulation (\textbf{31 Qubit Sim.} \cite{kechedzhi2023}), and a \textbf{TNO (SU)} \cite{Liao2023} approach using Heisenberg evolution and evolved with simple update, which is closely related to the BP approximation \cite{Alkabetz2021, TindallGauging2023}, and contracted exactly to compute expectation values.}
    \label{Fig:Z62FullComparison}
\end{figure*}
Here we compare data for the expectation value of $\langle Z_{62} \rangle$ after $20$ Trotter steps from a range of classical methods and the quantum processor. The results are shown in Fig. \ref{Fig:Z62FullComparison}. We include results at $\chi = 200$ and $\chi = 500$ from our BP-evolved TNS approach, as well as results from an extrapolation in $1/\chi$ to $\chi \rightarrow \infty$. The shaded region shows the difference between our finite $\chi = 500$ bond dimension data and the data extrapolated to infinite bond dimension, where we believe the true answer lies. For smaller bond dimensions the BP-evolved TNS method overshoots the true results in the region $\pi/4 \lesssim \theta_h \lesssim 7\pi/16$, an artifact which is resolved by extrapolating in the bond dimension. Our extrapolated BP-evolved TNS results are in close agreement with results obtained in \cite{begušić2023}, which uses a combination of a BP-evolved TNS and a BP-evolved tensor network operator (TNO) to perform a mixed Schr\"{o}dinger and Heisenberg evolution. Moreover, a method based on truncated Pauli strings \cite{begušić2023, Chan2023}, which approximates the Heisenberg evolution of the system and relies on a different type of approximation than tensor network methods and BP, also shows close agreement with our BP-evolved TNS approach and the aformentioned BP-evolved mixed TNS-TNO approach, although it slightly overshoots the value of $\langle Z_{62} \rangle$ for $\frac{5\pi}{32} \lesssim \theta_{h} \lesssim \frac{7\pi}{32}$ --- which can be observed in Fig. \ref{Fig:F2}d) where we have numerically accurate results from MPS calculations. 
\par Results in  Fig. \ref{Fig:Z62FullComparison} are also shown for a matrix product operator (MPO)-based method of simulating the Heisenberg evolution of the system \cite{Zalatel2023}. This approach, however, undershoots the value in the region $\frac{4\pi}{16} \lesssim \theta_{h} \lesssim \frac{5\pi}{16}$  --- mapping the system to a 1D ansatz suffers from the drawback of requiring much larger bond dimensions than the more general tensor network-based approaches. The limitations of methods based on 1D ansatzes for this problem, like MPS and MPO, is especially clear when considering the MPS data originally reported in \cite{kim2023}, along with the MPS results reported in this work in Fig. \ref{Fig:F2}. We also include data from calculations based on a tensor network operator (TNO) approach \cite{Liao2023}, extrapolated to $\chi \rightarrow \infty$, which is evolved using simple update (SU) --- which is closely related to BP \cite{Alkabetz2021, TindallGauging2023} --- and then is contracted exactly to compute expectation values\footnote{At least for TNS, we believe that BP contraction to compute expectations values is sufficiently accurate for this model based on comparisons to MPS in Fig. \ref{Fig:F2} and analysis of the validity of the BP approximation based on exact contraction and boundary MPS contraction in Fig. \ref{Fig:smallheavyhexdynamics} and Fig. \ref{Fig:BoundaryMPSMagnetisationComparison}.}. This method overshoots the true value of $\langle Z_{62} \rangle$ in the region $\frac{\pi}{8} \lesssim \theta_{h} \lesssim \frac{2\pi}{8}$  --- which can be directly observed in Fig. \ref{Fig:F2}d). Finally, we provide data from a smaller, $31$ qubit system calculation done in Ref. \cite{kechedzhi2023}.

\par We also would like to point out work in Ref. \cite{torre2023}, which used a dissipative mean-field theory to simulate the $127$-qubit system out to short circuit depths ($\sim 5$ Trotter steps) and obtained relatively accurate results --- although with some noticeable deviations from the true result, which can be computed to an accuracy $\sim 10^{-14}$ with our BP-evolved TNS method. 

\renewcommand{\theequation}{B\arabic{equation}}
\section*{Appendix B: MPS calculations}

\begin{figure*}[t!]
    \centering
    \includegraphics[width = \textwidth]{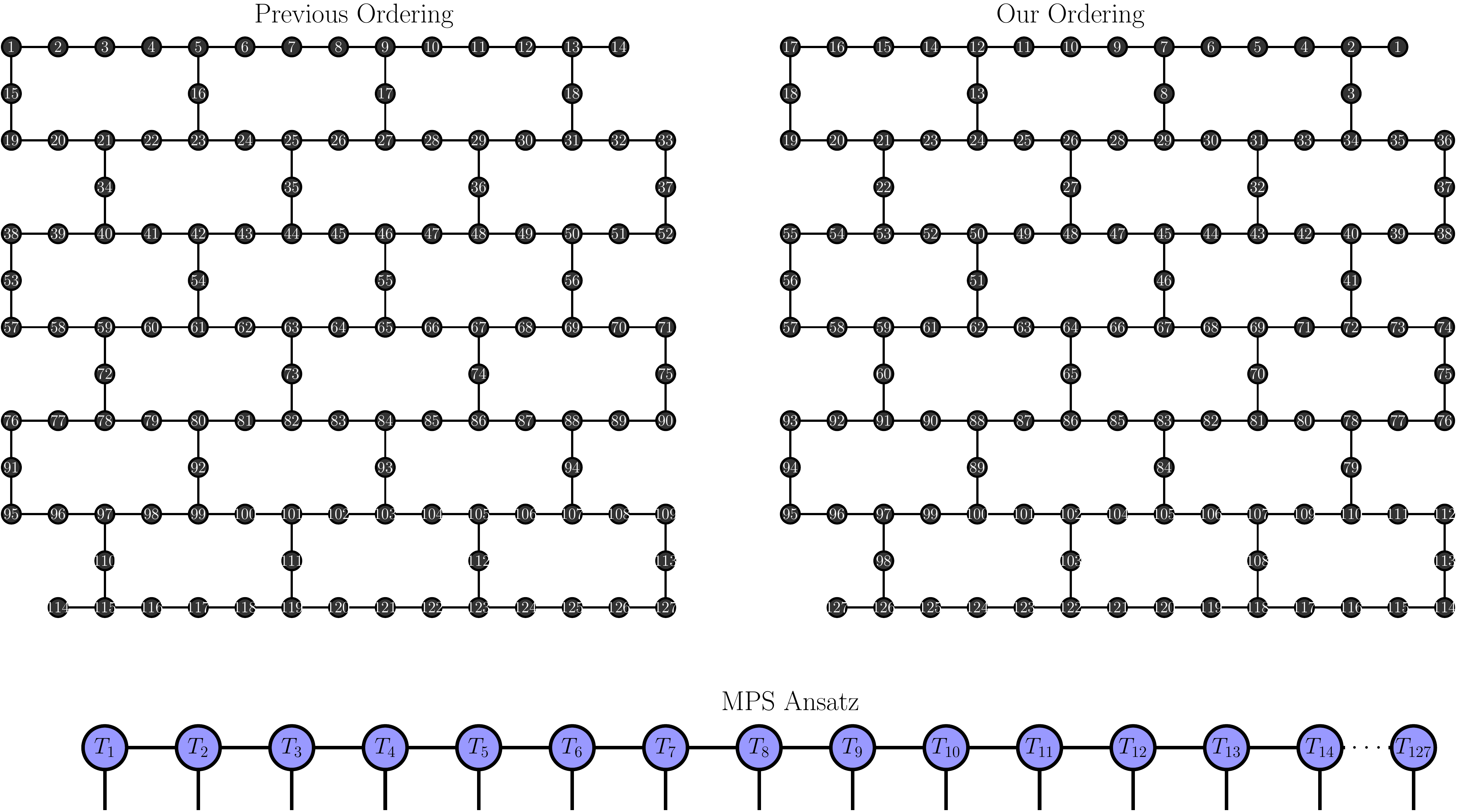}
    \caption{Top) Two possible mappings from the qubits of the Eagle processor to the sites of a MPS ansatz. The left diagram shows the ordering used in Ref. \cite{kim2023} while the right diagram shows the ordering we use to obtain the results in Fig. \ref{Fig:F2}d-f), which we find leads to higher accuracy for a given MPS bond dimension. Bottom) MPS ansatz with the site numbers corresponding to those in the above orderings.} 
    \label{Fig:MPSOrderings}
\end{figure*}

We performed MPS calculations for benchmarking in the unverifiable regime in Fig. \ref{Fig:F2}d-f). We implemented an ordering of the sites of the heavy-hex lattice which is distinct from that in Ref. \cite{kim2023}. Specifically, Fig. \ref{Fig:MPSOrderings} shows the qubits of the Eagle processor numbered as $1$ through $127$ which then map directly to the sites $1$ through $127$ of the MPS we use. Two numberings are shown: our own and the original ordering from Ref. \cite{kim2023}. We find that our own ordering results in slightly lower truncation errors when simulating the $127$-qubit kicked Ising model at fixed MPS bond dimension $\chi$. We break the two-site term in the propagator $U(\theta_{h})$ down into a product of, at most, $5$ commuting matrix product operators (MPOs) of bond dimension $2$. The MPS is then evolved by successively applying these MPOs and after each application truncation is performed of the state down to a maximum bond dimension $\chi$. The single-site gates are applied at the start of each Trotter step exactly. If implementing all two-site gates our decomposition of $U(\theta_{h})$ involves $5$ MPOs. In practice, however, we employ light cone depth reduction (LCDR), which means that if there are $n'$ Trotter steps left until we take a measurement on a specific site, we only include two-site gates in the MPOs which are within the remaining light cone of the simulation. This means that for a given $\theta_{h}$ in Fig. \ref{Fig:F2} we perform $20$ separate simulations to get the most accurate MPS results for the value of $\langle Z_{62} \rangle$ after each number of steps $n$.

In order to approximate the error of this simulation we calculate the sum of the singular values discarded during the application of a single MPO, which we call $\epsilon_{i}$, where $i$ refers to the MPO being applied. The total approximate error during the $n$ trotter step simulation (and plotted in the insets of Fig. \ref{Fig:F2}d-f) is then \cite{Zhou2020}
\begin{equation}
    E_{n} = 1 - \sum_{i=1}^{N}\left(1 - \epsilon_{i}\right)^{\frac{1}{N}},
    \label{Eq:ErrorApproximator}
\end{equation}
where the sum runs over all $N$ MPO-MPS applications (which may be less than $5n$ due to LCDR) during the simulation up to $n$ Trotter steps. The error per gate that we calculate is approximately the same as applying the gate exactly and taking the overlap with the truncated state.

\renewcommand{\theequation}{C\arabic{equation}}
\section*{Appendix C: Boundary MPS for the heavy-hex lattice}
\label{Subsec:BoundaryMPS}
\begin{figure*}[t!]
    \centering
    \includegraphics[width = \textwidth]{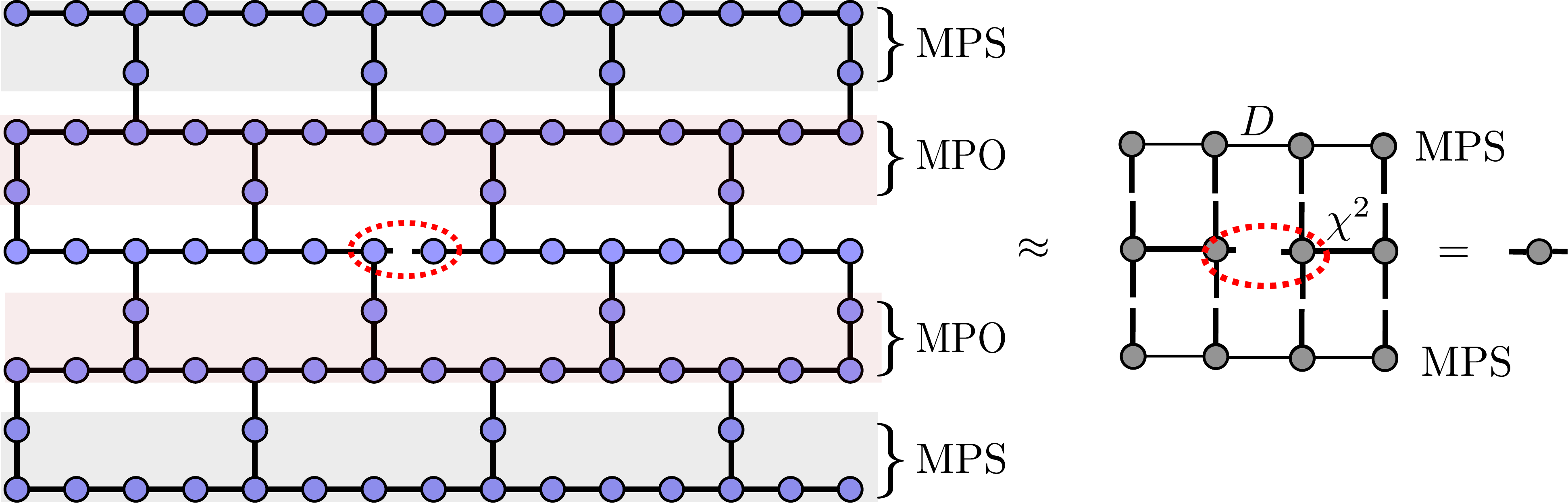}
    \caption{Approximating the edge environment of an edge $e$ of a TNS on the heavy-hex lattice of bond dimension $\chi$ using a boundary MPS-style scheme \cite{Ma2023}. The contraction of the norm network is done by successively (top to bottom and bottom to top) approximating the contraction of pairs of rows of the heavy hex lattice as MPSs of bond dimension $D$. The resulting contraction is then reduced to the contraction of a pair of MPSs incident to the given row of the TNS.} 
    \label{Fig:BoundaryMPSDiagram}
\end{figure*}

In order to approximate the error of BP using the separability of the edge environment (see the \hyperref[Subsec:BPErrorEstimate]{BP error estimate} section) for lattices that are too large to contract exactly, for example the larger lattices in Fig. \ref{Fig:smallheavyhexdynamics}d), we employ a method similar to boundary MPS \cite{Verstraete2004, Jordan2008, Lubasch2014} but generalized to arbitrary network structures \cite{Ma2023} to approximately contract the norm network down to a chosen edge to compute the edge environment. Specifically, we approximate the contraction of two rows of the heavy hex as an MPS of fixed bond dimension $D$ --- with the external indices of the given region of the TNS being mapped to the dangling indices of the MPS. This allows us to approximate the successive contraction of rows (from top to bottom and bottom to top) of the heavy-hex TNS as a sequence of MPS-MPO contractions.  This process is depicted diagramatically in Fig. \ref{Fig:BoundaryMPSDiagram}.

\begin{figure*}[t!]
    \centering
    \includegraphics[width = \textwidth]{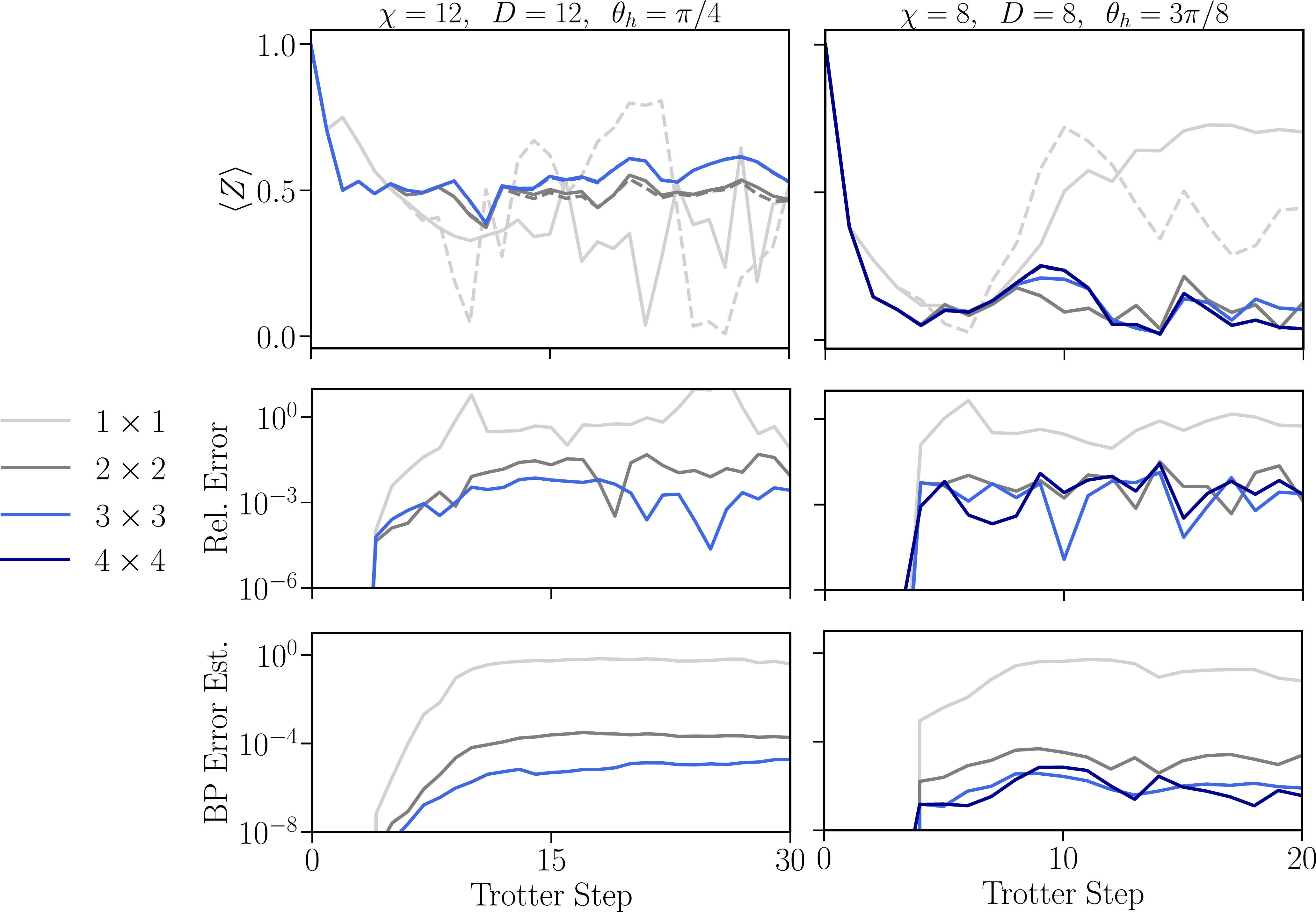}
    \caption{Dynamics of the single-site magnetization of the kicked Ising model on heavy-hex lattices of varying sizes. The TNS is evolved using the BP approximation and the magnetization is approximated with BP (solid lines) and boundary MPS (dashed lines) with the relevant parameters displayed above the plot. The middle plots show the relative error between calculating the expectation values with BP versus boundary MPS. Bottom plots show the BP error estimate based on Eq. (\ref{Eq:BPErrorEstimate}) for an edge incident to the site where the magnetization is calculated.} 
    \label{Fig:BoundaryMPSMagnetisationComparison}
\end{figure*}

We also computed the difference between calculating the magnetization of the BP-evolved TNS using the belief propagation method versus using boundary MPS for the parameters considered in Fig. \ref{Fig:smallheavyhexdynamics}d). We present the results here in Fig. \ref{Fig:BoundaryMPSMagnetisationComparison} and observe how close the values obtained are --- with the largest lattices showing differences between the two methods which are on the order of $10^{-4}$. The difference between the two methods generally decreases as the system size is increased.

\newpage

\bibliography{Bibliography}

\begin{thebibliography}{48}%
\makeatletter
\providecommand \@ifxundefined [1]{%
 \@ifx{#1\undefined}
}%
\providecommand \@ifnum [1]{%
 \ifnum #1\expandafter \@firstoftwo
 \else \expandafter \@secondoftwo
 \fi
}%
\providecommand \@ifx [1]{%
 \ifx #1\expandafter \@firstoftwo
 \else \expandafter \@secondoftwo
 \fi
}%
\providecommand \natexlab [1]{#1}%
\providecommand \enquote  [1]{``#1''}%
\providecommand \bibnamefont  [1]{#1}%
\providecommand \bibfnamefont [1]{#1}%
\providecommand \citenamefont [1]{#1}%
\providecommand \href@noop [0]{\@secondoftwo}%
\providecommand \href [0]{\begingroup \@sanitize@url \@href}%
\providecommand \@href[1]{\@@startlink{#1}\@@href}%
\providecommand \@@href[1]{\endgroup#1\@@endlink}%
\providecommand \@sanitize@url [0]{\catcode `\\12\catcode `\$12\catcode
  `\&12\catcode `\#12\catcode `\^12\catcode `\_12\catcode `\%12\relax}%
\providecommand \@@startlink[1]{}%
\providecommand \@@endlink[0]{}%
\providecommand \url  [0]{\begingroup\@sanitize@url \@url }%
\providecommand \@url [1]{\endgroup\@href {#1}{\urlprefix }}%
\providecommand \urlprefix  [0]{URL }%
\providecommand \Eprint [0]{\href }%
\providecommand \doibase [0]{https://doi.org/}%
\providecommand \selectlanguage [0]{\@gobble}%
\providecommand \bibinfo  [0]{\@secondoftwo}%
\providecommand \bibfield  [0]{\@secondoftwo}%
\providecommand \translation [1]{[#1]}%
\providecommand \BibitemOpen [0]{}%
\providecommand \bibitemStop [0]{}%
\providecommand \bibitemNoStop [0]{.\EOS\space}%
\providecommand \EOS [0]{\spacefactor3000\relax}%
\providecommand \BibitemShut  [1]{\csname bibitem#1\endcsname}%
\let\auto@bib@innerbib\@empty
\bibitem [{\citenamefont {Bravyi}\ \emph {et~al.}(2021)\citenamefont {Bravyi},
  \citenamefont {Gosset},\ and\ \citenamefont {Movassagh}}]{bravyi21}%
  \BibitemOpen
  \bibfield  {author} {\bibinfo {author} {\bibfnamefont {S.}~\bibnamefont
  {Bravyi}}, \bibinfo {author} {\bibfnamefont {D.}~\bibnamefont {Gosset}},\
  and\ \bibinfo {author} {\bibfnamefont {R.}~\bibnamefont {Movassagh}},\
  }\bibfield  {title} {\bibinfo {title} {Classical algorithms for quantum mean
  values},\ }\href {https://doi.org/10.1038/s41567-020-01109-8} {\bibfield
  {journal} {\bibinfo  {journal} {Nature Physics}\ }\textbf {\bibinfo {volume}
  {17}},\ \bibinfo {pages} {337} (\bibinfo {year} {2021})}\BibitemShut
  {NoStop}%
\bibitem [{\citenamefont {Wild}\ and\ \citenamefont {Alhambra}(2023)}]{wild23}%
  \BibitemOpen
  \bibfield  {author} {\bibinfo {author} {\bibfnamefont {D.~S.}\ \bibnamefont
  {Wild}}\ and\ \bibinfo {author} {\bibfnamefont {A.~M.}\ \bibnamefont
  {Alhambra}},\ }\bibfield  {title} {\bibinfo {title} {Classical simulation of
  short-time quantum dynamics},\ }\href
  {https://doi.org/10.1103/PRXQuantum.4.020340} {\bibfield  {journal} {\bibinfo
   {journal} {PRX Quantum}\ }\textbf {\bibinfo {volume} {4}},\ \bibinfo {pages}
  {020340} (\bibinfo {year} {2023})}\BibitemShut {NoStop}%
\bibitem [{\citenamefont {Aharonov}\ \emph {et~al.}(2023)\citenamefont
  {Aharonov}, \citenamefont {Gao}, \citenamefont {Landau}, \citenamefont
  {Liu},\ and\ \citenamefont {Vazirani}}]{aharonov23}%
  \BibitemOpen
  \bibfield  {author} {\bibinfo {author} {\bibfnamefont {D.}~\bibnamefont
  {Aharonov}}, \bibinfo {author} {\bibfnamefont {X.}~\bibnamefont {Gao}},
  \bibinfo {author} {\bibfnamefont {Z.}~\bibnamefont {Landau}}, \bibinfo
  {author} {\bibfnamefont {Y.}~\bibnamefont {Liu}},\ and\ \bibinfo {author}
  {\bibfnamefont {U.}~\bibnamefont {Vazirani}},\ }\bibfield  {title} {\bibinfo
  {title} {A polynomial-time classical algorithm for noisy random circuit
  sampling},\ }in\ \href {https://doi.org/10.1145/3564246.3585234} {\emph
  {\bibinfo {booktitle} {Proceedings of the 55th Annual ACM Symposium on Theory
  of Computing}}},\ \bibinfo {series and number} {STOC 2023}\ (\bibinfo
  {publisher} {Association for Computing Machinery},\ \bibinfo {address} {New
  York, NY, USA},\ \bibinfo {year} {2023})\ p.\ \bibinfo {pages}
  {945–957}\BibitemShut {NoStop}%
\bibitem [{\citenamefont {Aurate}\ \emph {et~al.}(2019)\citenamefont {Aurate}
  \emph {et~al.}}]{arute19}%
  \BibitemOpen
  \bibfield  {author} {\bibinfo {author} {\bibfnamefont {F.}~\bibnamefont
  {Aurate}} \emph {et~al.},\ }\bibfield  {title} {\bibinfo {title} {Quantum
  supremacy using a programmable superconducting processor},\ }\href
  {https://doi.org/10.1038/s41586-019-1666-5} {\bibfield  {journal} {\bibinfo
  {journal} {Nature}\ }\textbf {\bibinfo {volume} {574}},\ \bibinfo {pages}
  {505} (\bibinfo {year} {2019})}\BibitemShut {NoStop}%
\bibitem [{\citenamefont {Zhou}\ \emph {et~al.}(2020)\citenamefont {Zhou},
  \citenamefont {Stoudenmire},\ and\ \citenamefont {Waintal}}]{Zhou2020}%
  \BibitemOpen
  \bibfield  {author} {\bibinfo {author} {\bibfnamefont {Y.}~\bibnamefont
  {Zhou}}, \bibinfo {author} {\bibfnamefont {E.~M.}\ \bibnamefont
  {Stoudenmire}},\ and\ \bibinfo {author} {\bibfnamefont {X.}~\bibnamefont
  {Waintal}},\ }\bibfield  {title} {\bibinfo {title} {What {Limits} the
  {Simulation} of {Quantum} {Computers}?},\ }\href
  {https://doi.org/10.1103/PhysRevX.10.041038} {\bibfield  {journal} {\bibinfo
  {journal} {Physical Review X}\ }\textbf {\bibinfo {volume} {10}},\ \bibinfo
  {pages} {041038} (\bibinfo {year} {2020})}\BibitemShut {NoStop}%
\bibitem [{\citenamefont {Pan}\ and\ \citenamefont {Zhang}(2022)}]{Pan2022}%
  \BibitemOpen
  \bibfield  {author} {\bibinfo {author} {\bibfnamefont {F.}~\bibnamefont
  {Pan}}\ and\ \bibinfo {author} {\bibfnamefont {P.}~\bibnamefont {Zhang}},\
  }\bibfield  {title} {\bibinfo {title} {Simulation of {Quantum} {Circuits}
  {Using} the {Big}-{Batch} {Tensor} {Network} {Method}},\ }\href
  {https://doi.org/10.1103/PhysRevLett.128.030501} {\bibfield  {journal}
  {\bibinfo  {journal} {Physical Review Letters}\ }\textbf {\bibinfo {volume}
  {128}},\ \bibinfo {pages} {030501} (\bibinfo {year} {2022})}\BibitemShut
  {NoStop}%
\bibitem [{\citenamefont {Ayral}\ \emph {et~al.}(2023)\citenamefont {Ayral},
  \citenamefont {Louvet}, \citenamefont {Zhou}, \citenamefont {Lambert},
  \citenamefont {Stoudenmire},\ and\ \citenamefont {Waintal}}]{Ayral2023}%
  \BibitemOpen
  \bibfield  {author} {\bibinfo {author} {\bibfnamefont {T.}~\bibnamefont
  {Ayral}}, \bibinfo {author} {\bibfnamefont {T.}~\bibnamefont {Louvet}},
  \bibinfo {author} {\bibfnamefont {Y.}~\bibnamefont {Zhou}}, \bibinfo {author}
  {\bibfnamefont {C.}~\bibnamefont {Lambert}}, \bibinfo {author} {\bibfnamefont
  {E.~M.}\ \bibnamefont {Stoudenmire}},\ and\ \bibinfo {author} {\bibfnamefont
  {X.}~\bibnamefont {Waintal}},\ }\bibfield  {title} {\bibinfo {title}
  {Density-{Matrix} {Renormalization} {Group} {Algorithm} for {Simulating}
  {Quantum} {Circuits} with a {Finite} {Fidelity}},\ }\href
  {https://doi.org/10.1103/PRXQuantum.4.020304} {\bibfield  {journal} {\bibinfo
   {journal} {PRX Quantum}\ }\textbf {\bibinfo {volume} {4}},\ \bibinfo {pages}
  {020304} (\bibinfo {year} {2023})}\BibitemShut {NoStop}%
\bibitem [{\citenamefont {Temme}\ \emph {et~al.}(2017)\citenamefont {Temme},
  \citenamefont {Bravyi},\ and\ \citenamefont {Gambetta}}]{temme17}%
  \BibitemOpen
  \bibfield  {author} {\bibinfo {author} {\bibfnamefont {K.}~\bibnamefont
  {Temme}}, \bibinfo {author} {\bibfnamefont {S.}~\bibnamefont {Bravyi}},\ and\
  \bibinfo {author} {\bibfnamefont {J.~M.}\ \bibnamefont {Gambetta}},\
  }\bibfield  {title} {\bibinfo {title} {Error mitigation for short-depth
  quantum circuits},\ }\href {https://doi.org/10.1103/PhysRevLett.119.180509}
  {\bibfield  {journal} {\bibinfo  {journal} {Phys. Rev. Lett.}\ }\textbf
  {\bibinfo {volume} {119}},\ \bibinfo {pages} {180509} (\bibinfo {year}
  {2017})}\BibitemShut {NoStop}%
\bibitem [{\citenamefont {van~den Berg}\ \emph {et~al.}(2023)\citenamefont
  {van~den Berg}, \citenamefont {Minev}, \citenamefont {Kandala},\ and\
  \citenamefont {Temme}}]{vandenberg23}%
  \BibitemOpen
  \bibfield  {author} {\bibinfo {author} {\bibfnamefont {E.}~\bibnamefont
  {van~den Berg}}, \bibinfo {author} {\bibfnamefont {Z.~K.}\ \bibnamefont
  {Minev}}, \bibinfo {author} {\bibfnamefont {A.}~\bibnamefont {Kandala}},\
  and\ \bibinfo {author} {\bibfnamefont {K.}~\bibnamefont {Temme}},\ }\bibfield
   {title} {\bibinfo {title} {Probabilistic error cancellation with sparse
  {P}auli--{L}indblad models on noisy quantum processors},\ }\bibfield
  {journal} {\bibinfo  {journal} {Nature Physics}\ }\href
  {https://doi.org/10.1038/s41567-023-02042-2} {10.1038/s41567-023-02042-2}
  (\bibinfo {year} {2023})\BibitemShut {NoStop}%
\bibitem [{\citenamefont {Kim}\ \emph {et~al.}(2023)\citenamefont {Kim},
  \citenamefont {Eddins}, \citenamefont {Anand}, \citenamefont {Wei},
  \citenamefont {van~den Berg}, \citenamefont {Rosenblatt}, \citenamefont
  {Nayfeh}, \citenamefont {Wu}, \citenamefont {Zaletel}, \citenamefont
  {Temme},\ and\ \citenamefont {Kandala}}]{kim2023}%
  \BibitemOpen
  \bibfield  {author} {\bibinfo {author} {\bibfnamefont {Y.}~\bibnamefont
  {Kim}}, \bibinfo {author} {\bibfnamefont {A.}~\bibnamefont {Eddins}},
  \bibinfo {author} {\bibfnamefont {S.}~\bibnamefont {Anand}}, \bibinfo
  {author} {\bibfnamefont {K.~X.}\ \bibnamefont {Wei}}, \bibinfo {author}
  {\bibfnamefont {E.}~\bibnamefont {van~den Berg}}, \bibinfo {author}
  {\bibfnamefont {S.}~\bibnamefont {Rosenblatt}}, \bibinfo {author}
  {\bibfnamefont {H.}~\bibnamefont {Nayfeh}}, \bibinfo {author} {\bibfnamefont
  {Y.}~\bibnamefont {Wu}}, \bibinfo {author} {\bibfnamefont {M.}~\bibnamefont
  {Zaletel}}, \bibinfo {author} {\bibfnamefont {K.}~\bibnamefont {Temme}},\
  and\ \bibinfo {author} {\bibfnamefont {A.}~\bibnamefont {Kandala}},\
  }\bibfield  {title} {\bibinfo {title} {Evidence for the utility of quantum
  computing before fault tolerance},\ }\href
  {https://doi.org/10.1038/s41586-023-06096-3} {\bibfield  {journal} {\bibinfo
  {journal} {Nature}\ }\textbf {\bibinfo {volume} {618}},\ \bibinfo {pages}
  {500} (\bibinfo {year} {2023})}\BibitemShut {NoStop}%
\bibitem [{\citenamefont {Leifer}\ and\ \citenamefont
  {Poulin}(2008)}]{TensorNetworkBeliefPropagation1}%
  \BibitemOpen
  \bibfield  {author} {\bibinfo {author} {\bibfnamefont {M.}~\bibnamefont
  {Leifer}}\ and\ \bibinfo {author} {\bibfnamefont {D.}~\bibnamefont
  {Poulin}},\ }\bibfield  {title} {\bibinfo {title} {Quantum graphical models
  and belief propagation},\ }\href
  {https://doi.org/https://doi.org/10.1016/j.aop.2007.10.001} {\bibfield
  {journal} {\bibinfo  {journal} {Annals of Physics}\ }\textbf {\bibinfo
  {volume} {323}},\ \bibinfo {pages} {1899} (\bibinfo {year}
  {2008})}\BibitemShut {NoStop}%
\bibitem [{\citenamefont {Sahu}\ and\ \citenamefont
  {Swingle}(2022)}]{TensorNetworkBeliefPropagation2}%
  \BibitemOpen
  \bibfield  {author} {\bibinfo {author} {\bibfnamefont {S.}~\bibnamefont
  {Sahu}}\ and\ \bibinfo {author} {\bibfnamefont {B.}~\bibnamefont {Swingle}},\
  }\href@noop {} {\bibinfo {title} {Efficient tensor network simulation of
  quantum many-body physics on sparse graphs}} (\bibinfo {year} {2022}),\
  \Eprint {https://arxiv.org/abs/2206.04701} {arXiv:2206.04701 [quant-ph]}
  \BibitemShut {NoStop}%
\bibitem [{\citenamefont {Guo}\ \emph {et~al.}(2023)\citenamefont {Guo},
  \citenamefont {Poletti},\ and\ \citenamefont
  {Arad}}]{TensorNetworkBeliefPropagation3}%
  \BibitemOpen
  \bibfield  {author} {\bibinfo {author} {\bibfnamefont {C.}~\bibnamefont
  {Guo}}, \bibinfo {author} {\bibfnamefont {D.}~\bibnamefont {Poletti}},\ and\
  \bibinfo {author} {\bibfnamefont {I.}~\bibnamefont {Arad}},\ }\href@noop {}
  {\bibinfo {title} {Block belief propagation algorithm for 2d tensor
  networks}} (\bibinfo {year} {2023}),\ \Eprint
  {https://arxiv.org/abs/2301.05844} {arXiv:2301.05844 [quant-ph]} \BibitemShut
  {NoStop}%
\bibitem [{\citenamefont {Tindall}\ and\ \citenamefont
  {Fishman}(2023)}]{TindallGauging2023}%
  \BibitemOpen
  \bibfield  {author} {\bibinfo {author} {\bibfnamefont {J.}~\bibnamefont
  {Tindall}}\ and\ \bibinfo {author} {\bibfnamefont {M.}~\bibnamefont
  {Fishman}},\ }\href@noop {} {\bibinfo {title} {Gauging tensor networks with
  belief propagation}} (\bibinfo {year} {2023}),\ \Eprint
  {https://arxiv.org/abs/2306.17837} {arXiv:2306.17837 [quant-ph]} \BibitemShut
  {NoStop}%
\bibitem [{\citenamefont {Birnkammer}\ \emph {et~al.}(2022)\citenamefont
  {Birnkammer}, \citenamefont {Bastianello},\ and\ \citenamefont
  {Knap}}]{Birnkammer2022}%
  \BibitemOpen
  \bibfield  {author} {\bibinfo {author} {\bibfnamefont {S.}~\bibnamefont
  {Birnkammer}}, \bibinfo {author} {\bibfnamefont {A.}~\bibnamefont
  {Bastianello}},\ and\ \bibinfo {author} {\bibfnamefont {M.}~\bibnamefont
  {Knap}},\ }\bibfield  {title} {\bibinfo {title} {Prethermalization in
  one-dimensional quantum many-body systems with confinement},\ }\href
  {https://doi.org/10.1038/s41467-022-35301-6} {\bibfield  {journal} {\bibinfo
  {journal} {Nature Communications}\ }\textbf {\bibinfo {volume} {13}},\
  \bibinfo {pages} {7663} (\bibinfo {year} {2022})}\BibitemShut {NoStop}%
\bibitem [{\citenamefont {Kormos}\ \emph {et~al.}(2017)\citenamefont {Kormos},
  \citenamefont {Collura}, \citenamefont {Tak{\'a}cs},\ and\ \citenamefont
  {Calabrese}}]{Marton2017}%
  \BibitemOpen
  \bibfield  {author} {\bibinfo {author} {\bibfnamefont {M.}~\bibnamefont
  {Kormos}}, \bibinfo {author} {\bibfnamefont {M.}~\bibnamefont {Collura}},
  \bibinfo {author} {\bibfnamefont {G.}~\bibnamefont {Tak{\'a}cs}},\ and\
  \bibinfo {author} {\bibfnamefont {P.}~\bibnamefont {Calabrese}},\ }\bibfield
  {title} {\bibinfo {title} {Real-time confinement following a quantum quench
  to a non-integrable model},\ }\href {https://doi.org/10.1038/nphys3934}
  {\bibfield  {journal} {\bibinfo  {journal} {Nature Physics}\ }\textbf
  {\bibinfo {volume} {13}},\ \bibinfo {pages} {246} (\bibinfo {year}
  {2017})}\BibitemShut {NoStop}%
\bibitem [{\citenamefont {Liao}\ \emph {et~al.}(2023)\citenamefont {Liao},
  \citenamefont {Wang}, \citenamefont {Zhou}, \citenamefont {Zhang},\ and\
  \citenamefont {Xiang}}]{Liao2023}%
  \BibitemOpen
  \bibfield  {author} {\bibinfo {author} {\bibfnamefont {H.-J.}\ \bibnamefont
  {Liao}}, \bibinfo {author} {\bibfnamefont {K.}~\bibnamefont {Wang}}, \bibinfo
  {author} {\bibfnamefont {Z.-S.}\ \bibnamefont {Zhou}}, \bibinfo {author}
  {\bibfnamefont {P.}~\bibnamefont {Zhang}},\ and\ \bibinfo {author}
  {\bibfnamefont {T.}~\bibnamefont {Xiang}},\ }\href@noop {} {\bibinfo {title}
  {Simulation of {IBM}'s kicked ising experiment with {P}rojected {E}ntangled
  {P}air {O}perator}} (\bibinfo {year} {2023}),\ \Eprint
  {https://arxiv.org/abs/2308.03082} {arXiv:2308.03082 [quant-ph]} \BibitemShut
  {NoStop}%
\bibitem [{\citenamefont {Anand}\ \emph {et~al.}(2023)\citenamefont {Anand},
  \citenamefont {Temme}, \citenamefont {Kandala},\ and\ \citenamefont
  {Zaletel}}]{Zalatel2023}%
  \BibitemOpen
  \bibfield  {author} {\bibinfo {author} {\bibfnamefont {S.}~\bibnamefont
  {Anand}}, \bibinfo {author} {\bibfnamefont {K.}~\bibnamefont {Temme}},
  \bibinfo {author} {\bibfnamefont {A.}~\bibnamefont {Kandala}},\ and\ \bibinfo
  {author} {\bibfnamefont {M.}~\bibnamefont {Zaletel}},\ }\href@noop {}
  {\bibinfo {title} {Classical benchmarking of zero noise extrapolation beyond
  the exactly-verifiable regime}} (\bibinfo {year} {2023}),\ \Eprint
  {https://arxiv.org/abs/2306.17839} {arXiv:2306.17839 [quant-ph]} \BibitemShut
  {NoStop}%
\bibitem [{\citenamefont {Begušić}\ and\ \citenamefont
  {Chan}(2023)}]{Chan2023}%
  \BibitemOpen
  \bibfield  {author} {\bibinfo {author} {\bibfnamefont {T.}~\bibnamefont
  {Begušić}}\ and\ \bibinfo {author} {\bibfnamefont {G.~K.-L.}\ \bibnamefont
  {Chan}},\ }\href@noop {} {\bibinfo {title} {Fast classical simulation of
  evidence for the utility of quantum computing before fault tolerance}}
  (\bibinfo {year} {2023}),\ \Eprint {https://arxiv.org/abs/2306.16372}
  {arXiv:2306.16372 [quant-ph]} \BibitemShut {NoStop}%
\bibitem [{\citenamefont {Begušić}\ \emph {et~al.}(2023)\citenamefont
  {Begušić}, \citenamefont {Gray},\ and\ \citenamefont
  {Chan}}]{begušić2023}%
  \BibitemOpen
  \bibfield  {author} {\bibinfo {author} {\bibfnamefont {T.}~\bibnamefont
  {Begušić}}, \bibinfo {author} {\bibfnamefont {J.}~\bibnamefont {Gray}},\
  and\ \bibinfo {author} {\bibfnamefont {G.~K.-L.}\ \bibnamefont {Chan}},\
  }\href@noop {} {\bibinfo {title} {Fast and converged classical simulations of
  evidence for the utility of quantum computing before fault tolerance}}
  (\bibinfo {year} {2023}),\ \Eprint {https://arxiv.org/abs/2308.05077}
  {arXiv:2308.05077 [quant-ph]} \BibitemShut {NoStop}%
\bibitem [{\citenamefont {Kechedzhi}\ \emph {et~al.}(2023)\citenamefont
  {Kechedzhi}, \citenamefont {Isakov}, \citenamefont {Mandrà}, \citenamefont
  {Villalonga}, \citenamefont {Mi}, \citenamefont {Boixo},\ and\ \citenamefont
  {Smelyanskiy}}]{kechedzhi2023}%
  \BibitemOpen
  \bibfield  {author} {\bibinfo {author} {\bibfnamefont {K.}~\bibnamefont
  {Kechedzhi}}, \bibinfo {author} {\bibfnamefont {S.~V.}\ \bibnamefont
  {Isakov}}, \bibinfo {author} {\bibfnamefont {S.}~\bibnamefont {Mandrà}},
  \bibinfo {author} {\bibfnamefont {B.}~\bibnamefont {Villalonga}}, \bibinfo
  {author} {\bibfnamefont {X.}~\bibnamefont {Mi}}, \bibinfo {author}
  {\bibfnamefont {S.}~\bibnamefont {Boixo}},\ and\ \bibinfo {author}
  {\bibfnamefont {V.}~\bibnamefont {Smelyanskiy}},\ }\href@noop {} {\bibinfo
  {title} {Effective quantum volume, fidelity and computational cost of noisy
  quantum processing experiments}} (\bibinfo {year} {2023}),\ \Eprint
  {https://arxiv.org/abs/2306.15970} {arXiv:2306.15970 [quant-ph]} \BibitemShut
  {NoStop}%
\bibitem [{\citenamefont {Vlaar}\ and\ \citenamefont
  {Corboz}(2021)}]{Vlaar2021}%
  \BibitemOpen
  \bibfield  {author} {\bibinfo {author} {\bibfnamefont {P.~C.~G.}\
  \bibnamefont {Vlaar}}\ and\ \bibinfo {author} {\bibfnamefont
  {P.}~\bibnamefont {Corboz}},\ }\bibfield  {title} {\bibinfo {title}
  {Simulation of three-dimensional quantum systems with projected
  entangled-pair states},\ }\href {https://doi.org/10.1103/PhysRevB.103.205137}
  {\bibfield  {journal} {\bibinfo  {journal} {Physical Review B}\ }\textbf
  {\bibinfo {volume} {103}},\ \bibinfo {pages} {205137} (\bibinfo {year}
  {2021})}\BibitemShut {NoStop}%
\bibitem [{\citenamefont {Orús}\ and\ \citenamefont {Vidal}(2009)}]{Orus2009}%
  \BibitemOpen
  \bibfield  {author} {\bibinfo {author} {\bibfnamefont {R.}~\bibnamefont
  {Orús}}\ and\ \bibinfo {author} {\bibfnamefont {G.}~\bibnamefont {Vidal}},\
  }\bibfield  {title} {\bibinfo {title} {Simulation of two-dimensional quantum
  systems on an infinite lattice revisited: {Corner} transfer matrix for tensor
  contraction},\ }\href {https://doi.org/10.1103/PhysRevB.80.094403} {\bibfield
   {journal} {\bibinfo  {journal} {Physical Review B}\ }\textbf {\bibinfo
  {volume} {80}},\ \bibinfo {pages} {094403} (\bibinfo {year}
  {2009})}\BibitemShut {NoStop}%
\bibitem [{\citenamefont {Phien}\ \emph
  {et~al.}(2015{\natexlab{a}})\citenamefont {Phien}, \citenamefont {Bengua},
  \citenamefont {Tuan}, \citenamefont {Corboz},\ and\ \citenamefont
  {Orús}}]{Phien2015a}%
  \BibitemOpen
  \bibfield  {author} {\bibinfo {author} {\bibfnamefont {H.~N.}\ \bibnamefont
  {Phien}}, \bibinfo {author} {\bibfnamefont {J.~A.}\ \bibnamefont {Bengua}},
  \bibinfo {author} {\bibfnamefont {H.~D.}\ \bibnamefont {Tuan}}, \bibinfo
  {author} {\bibfnamefont {P.}~\bibnamefont {Corboz}},\ and\ \bibinfo {author}
  {\bibfnamefont {R.}~\bibnamefont {Orús}},\ }\bibfield  {title} {\bibinfo
  {title} {Infinite projected entangled pair states algorithm improved: {Fast}
  full update and gauge fixing},\ }\href
  {https://doi.org/10.1103/PhysRevB.92.035142} {\bibfield  {journal} {\bibinfo
  {journal} {Physical Review B}\ }\textbf {\bibinfo {volume} {92}},\ \bibinfo
  {pages} {035142} (\bibinfo {year} {2015}{\natexlab{a}})}\BibitemShut
  {NoStop}%
\bibitem [{\citenamefont {Jiang}\ \emph {et~al.}(2008)\citenamefont {Jiang},
  \citenamefont {Weng},\ and\ \citenamefont {Xiang}}]{Jiang2008}%
  \BibitemOpen
  \bibfield  {author} {\bibinfo {author} {\bibfnamefont {H.~C.}\ \bibnamefont
  {Jiang}}, \bibinfo {author} {\bibfnamefont {Z.~Y.}\ \bibnamefont {Weng}},\
  and\ \bibinfo {author} {\bibfnamefont {T.}~\bibnamefont {Xiang}},\ }\bibfield
   {title} {\bibinfo {title} {Accurate {Determination} of {Tensor} {Network}
  {State} of {Quantum} {Lattice} {Models} in {Two} {Dimensions}},\ }\href
  {https://doi.org/10.1103/PhysRevLett.101.090603} {\bibfield  {journal}
  {\bibinfo  {journal} {Physical Review Letters}\ }\textbf {\bibinfo {volume}
  {101}},\ \bibinfo {pages} {090603} (\bibinfo {year} {2008})}\BibitemShut
  {NoStop}%
\bibitem [{\citenamefont {Vidal}(2003)}]{Vidal2003}%
  \BibitemOpen
  \bibfield  {author} {\bibinfo {author} {\bibfnamefont {G.}~\bibnamefont
  {Vidal}},\ }\bibfield  {title} {\bibinfo {title} {Efficient {Classical}
  {Simulation} of {Slightly} {Entangled} {Quantum} {Computations}},\ }\href
  {https://doi.org/10.1103/PhysRevLett.91.147902} {\bibfield  {journal}
  {\bibinfo  {journal} {Physical Review Letters}\ }\textbf {\bibinfo {volume}
  {91}},\ \bibinfo {pages} {147902} (\bibinfo {year} {2003})}\BibitemShut
  {NoStop}%
\bibitem [{\citenamefont {Jahromi}\ and\ \citenamefont
  {Orús}(2019)}]{Jahromi2019}%
  \BibitemOpen
  \bibfield  {author} {\bibinfo {author} {\bibfnamefont {S.~S.}\ \bibnamefont
  {Jahromi}}\ and\ \bibinfo {author} {\bibfnamefont {R.}~\bibnamefont
  {Orús}},\ }\bibfield  {title} {\bibinfo {title} {Universal tensor-network
  algorithm for any infinite lattice},\ }\href
  {https://doi.org/10.1103/PhysRevB.99.195105} {\bibfield  {journal} {\bibinfo
  {journal} {Physical Review B}\ }\textbf {\bibinfo {volume} {99}},\ \bibinfo
  {pages} {195105} (\bibinfo {year} {2019})}\BibitemShut {NoStop}%
\bibitem [{\citenamefont {Ran}\ \emph {et~al.}(2012)\citenamefont {Ran},
  \citenamefont {Li}, \citenamefont {Xi}, \citenamefont {Zhang},\ and\
  \citenamefont {Su}}]{Ran2012}%
  \BibitemOpen
  \bibfield  {author} {\bibinfo {author} {\bibfnamefont {S.-J.}\ \bibnamefont
  {Ran}}, \bibinfo {author} {\bibfnamefont {W.}~\bibnamefont {Li}}, \bibinfo
  {author} {\bibfnamefont {B.}~\bibnamefont {Xi}}, \bibinfo {author}
  {\bibfnamefont {Z.}~\bibnamefont {Zhang}},\ and\ \bibinfo {author}
  {\bibfnamefont {G.}~\bibnamefont {Su}},\ }\bibfield  {title} {\bibinfo
  {title} {Optimized decimation of tensor networks with super-orthogonalization
  for two-dimensional quantum lattice models},\ }\href
  {https://doi.org/10.1103/PhysRevB.86.134429} {\bibfield  {journal} {\bibinfo
  {journal} {Physical Review B}\ }\textbf {\bibinfo {volume} {86}},\ \bibinfo
  {pages} {134429} (\bibinfo {year} {2012})}\BibitemShut {NoStop}%
\bibitem [{\citenamefont {Phien}\ \emph
  {et~al.}(2015{\natexlab{b}})\citenamefont {Phien}, \citenamefont
  {McCulloch},\ and\ \citenamefont {Vidal}}]{Phien2015}%
  \BibitemOpen
  \bibfield  {author} {\bibinfo {author} {\bibfnamefont {H.~N.}\ \bibnamefont
  {Phien}}, \bibinfo {author} {\bibfnamefont {I.~P.}\ \bibnamefont
  {McCulloch}},\ and\ \bibinfo {author} {\bibfnamefont {G.}~\bibnamefont
  {Vidal}},\ }\bibfield  {title} {\bibinfo {title} {Fast convergence of
  imaginary time evolution tensor network algorithms by recycling the
  environment},\ }\href {https://doi.org/10.1103/PhysRevB.91.115137} {\bibfield
   {journal} {\bibinfo  {journal} {Physical Review B}\ }\textbf {\bibinfo
  {volume} {91}},\ \bibinfo {pages} {115137} (\bibinfo {year}
  {2015}{\natexlab{b}})}\BibitemShut {NoStop}%
\bibitem [{\citenamefont {Alkabetz}\ and\ \citenamefont
  {Arad}(2021{\natexlab{a}})}]{SimpleUpdateGauging3}%
  \BibitemOpen
  \bibfield  {author} {\bibinfo {author} {\bibfnamefont {R.}~\bibnamefont
  {Alkabetz}}\ and\ \bibinfo {author} {\bibfnamefont {I.}~\bibnamefont
  {Arad}},\ }\bibfield  {title} {\bibinfo {title} {Tensor networks contraction
  and the belief propagation algorithm},\ }\href
  {https://doi.org/10.1103/PhysRevResearch.3.023073} {\bibfield  {journal}
  {\bibinfo  {journal} {Phys. Rev. Res.}\ }\textbf {\bibinfo {volume} {3}},\
  \bibinfo {pages} {023073} (\bibinfo {year} {2021}{\natexlab{a}})}\BibitemShut
  {NoStop}%
\bibitem [{\citenamefont {Verstraete}\ and\ \citenamefont
  {Cirac}(2004)}]{Verstraete2004}%
  \BibitemOpen
  \bibfield  {author} {\bibinfo {author} {\bibfnamefont {F.}~\bibnamefont
  {Verstraete}}\ and\ \bibinfo {author} {\bibfnamefont {J.~I.}\ \bibnamefont
  {Cirac}},\ }\href {https://doi.org/10.48550/arXiv.cond-mat/0407066} {\emph
  {\bibinfo {title} {Renormalization algorithms for {Quantum}-{Many} {Body}
  {Systems} in two and higher dimensions}}},\ \bibinfo {type} {Tech. Rep.}\
  (\bibinfo {year} {2004})\ \bibinfo {note} {arXiv:cond-mat/0407066 type:
  article}\BibitemShut {NoStop}%
\bibitem [{\citenamefont {Tindall}\ \emph {et~al.}(2022)\citenamefont
  {Tindall}, \citenamefont {Searle}, \citenamefont {Alhajri},\ and\
  \citenamefont {Jaksch}}]{tindall2022quantum}%
  \BibitemOpen
  \bibfield  {author} {\bibinfo {author} {\bibfnamefont {J.}~\bibnamefont
  {Tindall}}, \bibinfo {author} {\bibfnamefont {A.}~\bibnamefont {Searle}},
  \bibinfo {author} {\bibfnamefont {A.}~\bibnamefont {Alhajri}},\ and\ \bibinfo
  {author} {\bibfnamefont {D.}~\bibnamefont {Jaksch}},\ }\bibfield  {title}
  {\bibinfo {title} {Quantum physics in connected worlds},\ }\href@noop {}
  {\bibfield  {journal} {\bibinfo  {journal} {Nature Communications}\ }\textbf
  {\bibinfo {volume} {13}},\ \bibinfo {pages} {7445} (\bibinfo {year}
  {2022})}\BibitemShut {NoStop}%
\bibitem [{\citenamefont {Searle}\ and\ \citenamefont
  {Tindall}(2023)}]{searle2023exact}%
  \BibitemOpen
  \bibfield  {author} {\bibinfo {author} {\bibfnamefont {A.}~\bibnamefont
  {Searle}}\ and\ \bibinfo {author} {\bibfnamefont {J.}~\bibnamefont
  {Tindall}},\ }\href@noop {} {\bibinfo {title} {An exact continuous theory for
  spin systems on graphons}} (\bibinfo {year} {2023}),\ \Eprint
  {https://arxiv.org/abs/2303.00731} {arXiv:2303.00731 [cond-mat.stat-mech]}
  \BibitemShut {NoStop}%
\bibitem [{ITe(2023)}]{ITensorNetworks}%
  \BibitemOpen
  \href@noop {} {\bibinfo {title} {{ITensorNetworks.jl}}},\ \bibinfo
  {howpublished} {\url{https://github.com/mtfishman/ITensorNetworks.jl}}
  (\bibinfo {year} {2023})\BibitemShut {NoStop}%
\bibitem [{\citenamefont {Wang}\ \emph {et~al.}(2011)\citenamefont {Wang},
  \citenamefont {Pižorn},\ and\ \citenamefont {Verstraete}}]{Wang2011}%
  \BibitemOpen
  \bibfield  {author} {\bibinfo {author} {\bibfnamefont {L.}~\bibnamefont
  {Wang}}, \bibinfo {author} {\bibfnamefont {I.}~\bibnamefont {Pižorn}},\ and\
  \bibinfo {author} {\bibfnamefont {F.}~\bibnamefont {Verstraete}},\ }\bibfield
   {title} {\bibinfo {title} {Monte {Carlo} simulation with tensor network
  states},\ }\href {https://doi.org/10.1103/PhysRevB.83.134421} {\bibfield
  {journal} {\bibinfo  {journal} {Physical Review B}\ }\textbf {\bibinfo
  {volume} {83}},\ \bibinfo {pages} {134421} (\bibinfo {year}
  {2011})}\BibitemShut {NoStop}%
\bibitem [{\citenamefont {Li}\ \emph {et~al.}(2012)\citenamefont {Li},
  \citenamefont {von Delft},\ and\ \citenamefont {Xiang}}]{Li2012}%
  \BibitemOpen
  \bibfield  {author} {\bibinfo {author} {\bibfnamefont {W.}~\bibnamefont
  {Li}}, \bibinfo {author} {\bibfnamefont {J.}~\bibnamefont {von Delft}},\ and\
  \bibinfo {author} {\bibfnamefont {T.}~\bibnamefont {Xiang}},\ }\bibfield
  {title} {\bibinfo {title} {Efficient simulation of infinite tree tensor
  network states on the {Bethe} lattice},\ }\href
  {https://doi.org/10.1103/PhysRevB.86.195137} {\bibfield  {journal} {\bibinfo
  {journal} {Physical Review B}\ }\textbf {\bibinfo {volume} {86}},\ \bibinfo
  {pages} {195137} (\bibinfo {year} {2012})}\BibitemShut {NoStop}%
\bibitem [{\citenamefont {Alkabetz}\ and\ \citenamefont
  {Arad}(2021{\natexlab{b}})}]{Alkabetz2021}%
  \BibitemOpen
  \bibfield  {author} {\bibinfo {author} {\bibfnamefont {R.}~\bibnamefont
  {Alkabetz}}\ and\ \bibinfo {author} {\bibfnamefont {I.}~\bibnamefont
  {Arad}},\ }\bibfield  {title} {\bibinfo {title} {Tensor networks contraction
  and the belief propagation algorithm},\ }\href
  {https://doi.org/10.1103/PhysRevResearch.3.023073} {\bibfield  {journal}
  {\bibinfo  {journal} {Physical Review Research}\ }\textbf {\bibinfo {volume}
  {3}},\ \bibinfo {pages} {023073} (\bibinfo {year}
  {2021}{\natexlab{b}})}\BibitemShut {NoStop}%
\bibitem [{\citenamefont {Pearl}(1982)}]{BeliefPropagation1}%
  \BibitemOpen
  \bibfield  {author} {\bibinfo {author} {\bibfnamefont {J.}~\bibnamefont
  {Pearl}},\ }\bibfield  {title} {\bibinfo {title} {Reverend bayes on inference
  engines: A distributed hierarchical approach},\ }in\ \href@noop {} {\emph
  {\bibinfo {booktitle} {Proceedings of the Second AAAI Conference on
  Artificial Intelligence}}},\ \bibinfo {series and number} {AAAI'82}\
  (\bibinfo  {publisher} {AAAI Press},\ \bibinfo {year} {1982})\ p.\ \bibinfo
  {pages} {133–136}\BibitemShut {NoStop}%
\bibitem [{\citenamefont {Pippan}\ \emph {et~al.}(2010)\citenamefont {Pippan},
  \citenamefont {White},\ and\ \citenamefont {Evertz}}]{Pippan2010}%
  \BibitemOpen
  \bibfield  {author} {\bibinfo {author} {\bibfnamefont {P.}~\bibnamefont
  {Pippan}}, \bibinfo {author} {\bibfnamefont {S.~R.}\ \bibnamefont {White}},\
  and\ \bibinfo {author} {\bibfnamefont {H.~G.}\ \bibnamefont {Evertz}},\
  }\bibfield  {title} {\bibinfo {title} {Efficient matrix-product state method
  for periodic boundary conditions},\ }\href
  {https://doi.org/10.1103/PhysRevB.81.081103} {\bibfield  {journal} {\bibinfo
  {journal} {Physical Review B}\ }\textbf {\bibinfo {volume} {81}},\ \bibinfo
  {pages} {081103} (\bibinfo {year} {2010})}\BibitemShut {NoStop}%
\bibitem [{\citenamefont {Evenbly}(2018)}]{Evenbly2018}%
  \BibitemOpen
  \bibfield  {author} {\bibinfo {author} {\bibfnamefont {G.}~\bibnamefont
  {Evenbly}},\ }\bibfield  {title} {\bibinfo {title} {Gauge fixing, canonical
  forms, and optimal truncations in tensor networks with closed loops},\ }\href
  {https://doi.org/10.1103/PhysRevB.98.085155} {\bibfield  {journal} {\bibinfo
  {journal} {Physical Review B}\ }\textbf {\bibinfo {volume} {98}},\ \bibinfo
  {pages} {085155} (\bibinfo {year} {2018})}\BibitemShut {NoStop}%
\bibitem [{\citenamefont {Depireux}\ \emph {et~al.}(2001)\citenamefont
  {Depireux}, \citenamefont {Simon}, \citenamefont {Klein},\ and\ \citenamefont
  {Shamma}}]{Shihab2001}%
  \BibitemOpen
  \bibfield  {author} {\bibinfo {author} {\bibfnamefont {D.~A.}\ \bibnamefont
  {Depireux}}, \bibinfo {author} {\bibfnamefont {J.~Z.}\ \bibnamefont {Simon}},
  \bibinfo {author} {\bibfnamefont {D.~J.}\ \bibnamefont {Klein}},\ and\
  \bibinfo {author} {\bibfnamefont {S.~A.}\ \bibnamefont {Shamma}},\ }\bibfield
   {title} {\bibinfo {title} {Spectro-temporal response field characterization
  with dynamic ripples in ferret primary auditory cortex},\ }\href
  {https://doi.org/10.1152/jn.2001.85.3.1220} {\bibfield  {journal} {\bibinfo
  {journal} {Journal of Neurophysiology}\ }\textbf {\bibinfo {volume} {85}},\
  \bibinfo {pages} {1220} (\bibinfo {year} {2001})},\ \bibinfo {note} {pMID:
  11247991},\ \Eprint
  {https://arxiv.org/abs/https://doi.org/10.1152/jn.2001.85.3.1220}
  {https://doi.org/10.1152/jn.2001.85.3.1220} \BibitemShut {NoStop}%
\bibitem [{\citenamefont {Bezanson}\ \emph {et~al.}(2017)\citenamefont
  {Bezanson}, \citenamefont {Edelman}, \citenamefont {Karpinski},\ and\
  \citenamefont {Shah}}]{bezanson2017julia}%
  \BibitemOpen
  \bibfield  {author} {\bibinfo {author} {\bibfnamefont {J.}~\bibnamefont
  {Bezanson}}, \bibinfo {author} {\bibfnamefont {A.}~\bibnamefont {Edelman}},
  \bibinfo {author} {\bibfnamefont {S.}~\bibnamefont {Karpinski}},\ and\
  \bibinfo {author} {\bibfnamefont {V.~B.}\ \bibnamefont {Shah}},\ }\bibfield
  {title} {\bibinfo {title} {Julia: A fresh approach to numerical computing},\
  }\href {https://doi.org/https://doi.org/10.1137/141000671} {\bibfield
  {journal} {\bibinfo  {journal} {SIAM review}\ }\textbf {\bibinfo {volume}
  {59}},\ \bibinfo {pages} {65} (\bibinfo {year} {2017})}\BibitemShut {NoStop}%
\bibitem [{\citenamefont {Fishman}\ \emph {et~al.}(2022)\citenamefont
  {Fishman}, \citenamefont {White},\ and\ \citenamefont
  {Stoudenmire}}]{itensor-r0.3}%
  \BibitemOpen
  \bibfield  {author} {\bibinfo {author} {\bibfnamefont {M.}~\bibnamefont
  {Fishman}}, \bibinfo {author} {\bibfnamefont {S.~R.}\ \bibnamefont {White}},\
  and\ \bibinfo {author} {\bibfnamefont {E.~M.}\ \bibnamefont {Stoudenmire}},\
  }\bibfield  {title} {\bibinfo {title} {{Codebase release 0.3 for ITensor}},\
  }\href {https://doi.org/10.21468/SciPostPhysCodeb.4-r0.3} {\bibfield
  {journal} {\bibinfo  {journal} {SciPost Phys. Codebases}\ ,\ \bibinfo {pages}
  {4}} (\bibinfo {year} {2022})}\BibitemShut {NoStop}%
\bibitem [{Gra(2023)}]{GraphTikz}%
  \BibitemOpen
  \href@noop {} {\bibinfo {title} {{GraphTikz.jl}}},\ \bibinfo {howpublished}
  {\url{https://github.com/mtfishman/GraphTikZ.jl}} (\bibinfo {year}
  {2023})\BibitemShut {NoStop}%
\bibitem [{\citenamefont {Torre}\ and\ \citenamefont
  {Roses}(2023)}]{torre2023}%
  \BibitemOpen
  \bibfield  {author} {\bibinfo {author} {\bibfnamefont {E.~G.~D.}\
  \bibnamefont {Torre}}\ and\ \bibinfo {author} {\bibfnamefont {M.~M.}\
  \bibnamefont {Roses}},\ }\href@noop {} {\bibinfo {title} {Dissipative
  mean-field theory of ibm utility experiment}} (\bibinfo {year} {2023}),\
  \Eprint {https://arxiv.org/abs/2308.01339} {arXiv:2308.01339 [quant-ph]}
  \BibitemShut {NoStop}%
\bibitem [{\citenamefont {Jordan}\ \emph {et~al.}(2008)\citenamefont {Jordan},
  \citenamefont {Orús}, \citenamefont {Vidal}, \citenamefont {Verstraete},\
  and\ \citenamefont {Cirac}}]{Jordan2008}%
  \BibitemOpen
  \bibfield  {author} {\bibinfo {author} {\bibfnamefont {J.}~\bibnamefont
  {Jordan}}, \bibinfo {author} {\bibfnamefont {R.}~\bibnamefont {Orús}},
  \bibinfo {author} {\bibfnamefont {G.}~\bibnamefont {Vidal}}, \bibinfo
  {author} {\bibfnamefont {F.}~\bibnamefont {Verstraete}},\ and\ \bibinfo
  {author} {\bibfnamefont {J.~I.}\ \bibnamefont {Cirac}},\ }\bibfield  {title}
  {\bibinfo {title} {Classical {Simulation} of {Infinite}-{Size} {Quantum}
  {Lattice} {Systems} in {Two} {Spatial} {Dimensions}},\ }\href
  {https://doi.org/10.1103/PhysRevLett.101.250602} {\bibfield  {journal}
  {\bibinfo  {journal} {Physical Review Letters}\ }\textbf {\bibinfo {volume}
  {101}},\ \bibinfo {pages} {250602} (\bibinfo {year} {2008})}\BibitemShut
  {NoStop}%
\bibitem [{\citenamefont {Lubasch}\ \emph {et~al.}(2014)\citenamefont
  {Lubasch}, \citenamefont {Cirac},\ and\ \citenamefont
  {Bañuls}}]{Lubasch2014}%
  \BibitemOpen
  \bibfield  {author} {\bibinfo {author} {\bibfnamefont {M.}~\bibnamefont
  {Lubasch}}, \bibinfo {author} {\bibfnamefont {J.~I.}\ \bibnamefont {Cirac}},\
  and\ \bibinfo {author} {\bibfnamefont {M.-C.}\ \bibnamefont {Bañuls}},\
  }\bibfield  {title} {\bibinfo {title} {Unifying projected entangled pair
  state contractions},\ }\href {https://doi.org/10.1088/1367-2630/16/3/033014}
  {\bibfield  {journal} {\bibinfo  {journal} {New Journal of Physics}\ }\textbf
  {\bibinfo {volume} {16}},\ \bibinfo {pages} {033014} (\bibinfo {year}
  {2014})}\BibitemShut {NoStop}%
\bibitem [{\citenamefont {Ma}\ \emph {et~al.}(2023)\citenamefont {Ma},
  \citenamefont {Fishman}, \citenamefont {Stoudenmire},\ and\ \citenamefont
  {Solomonik}}]{Ma2023}%
  \BibitemOpen
  \bibfield  {author} {\bibinfo {author} {\bibfnamefont {L.}~\bibnamefont
  {Ma}}, \bibinfo {author} {\bibfnamefont {M.}~\bibnamefont {Fishman}},
  \bibinfo {author} {\bibfnamefont {E.~M.}\ \bibnamefont {Stoudenmire}},\ and\
  \bibinfo {author} {\bibfnamefont {E.}~\bibnamefont {Solomonik}},\ }\href@noop
  {} {\bibinfo {title} {\emph{In preparation}}} (\bibinfo {year}
  {2023})\BibitemShut {NoStop}%
\end{thebibliography}%

\end{document}